\documentclass[aps,pra, twocolumn, superscriptaddress, floatfix]{revtex4-1} 

\usepackage[pdftex]{graphicx} 
\usepackage{amsfonts,amssymb,amsbsy, amsmath}  
\usepackage[separate-uncertainty=true]{siunitx}
\usepackage{caption}
\usepackage{subcaption}
\usepackage{color}
\usepackage{multirow}
\usepackage{hyperref}
\usepackage{varioref}
\usepackage{cleveref}
\usepackage{ulem}
\usepackage{afterpage}
\graphicspath{{./figures/}}

\begin{document}

\newcommand{\filip}[1]{\textcolor{blue}{{\bf Filip: #1 }}}
\newcommand{\antti}[1]{\textcolor{green}{{\bf Antti: #1 }}}
\newcommand{\ilja}[1]{\textcolor{red}{{\bf Ilja: #1 }}}
\newcommand{\gao}{$\beta$-Ga$_2$O$_3$}
\newcommand{\via}{$V_{\mathrm{Ga}}^{\mathrm{ia}}$}
\newcommand{\vib}{$V_{\mathrm{Ga}}^{\mathrm{ib}}$}
\newcommand{\vic}{$V_{\mathrm{Ga}}^{\mathrm{ic}}$}
\newcommand{\vgai}{$V_{\mathrm{Ga1}}$}
\newcommand{\vgaii}{$V_{\mathrm{Ga2}}$}
\newcommand{\taub}{$\tau_{\mathrm{B}}$}

\title{Split Ga vacancies and the unusually strong anisotropy of positron annihilation spectra in $\boldsymbol{\beta}$-Ga$_2$O$_3$}

\author{Antti Karjalainen}
\affiliation{Department of Physics, University of Helsinki, P.O. Box 43, FI-00014 Helsinki, Finland}
\affiliation{Department of Applied Physics, Aalto University, P.O. Box 15100, FI-00076 Espoo, Finland}
\author{Vera Prozheeva}
\affiliation{Department of Applied Physics, Aalto University, P.O. Box 15100, FI-00076 Espoo, Finland}
\author{Kristoffer Simula}
\affiliation{Department of Physics, University of Helsinki, P.O. Box 43, FI-00014 Helsinki, Finland}
\author{Ilja Makkonen}
\affiliation{Department of Physics, University of Helsinki, P.O. Box 43, FI-00014 Helsinki, Finland}
\affiliation{Department of Applied Physics, Aalto University, P.O. Box 15100, FI-00076 Espoo, Finland}
\affiliation{Helsinki Institute of Physics, P.O. Box 64, FI-00014 Helsinki, Finland}
\author{Vincent Callewaert}
\affiliation{Department of Physics, Universiteit Antwerpen, Antwerpen 2020, Belgium}
\author{Joel B. Varley}
\affiliation{Lawrence Livermore National Laboratory, Livermore, CA 94550, USA}
\author{Filip Tuomisto}
\affiliation{Department of Physics, University of Helsinki, P.O. Box 43, FI-00014 Helsinki, Finland}
\affiliation{Department of Applied Physics, Aalto University, P.O. Box 15100, FI-00076 Espoo, Finland}
\affiliation{Helsinki Institute of Physics, P.O. Box 64, FI-00014 Helsinki, Finland}

\date{\today}

\begin{abstract}

We report a systematic first principles study on positron annihilation parameters in the $\beta$-Ga$_2$O$_3$ lattice and Ga mono-vacancy defects complemented with orientation-dependent experiments of the Doppler broadening of the positron-electron annihilation. We find that both the $\beta$-Ga$_2$O$_3$ lattice and the considered defects exhibit unusually strong anisotropy in their Doppler broadening signals. This anisotropy is associated with low symmetry of the $\beta$-Ga$_2$O$_3$ crystal structure that leads to unusual kind of one-dimensional confinement of positrons even in the delocalized state in the lattice. In particular, the split Ga vacancies recently observed by scanning transmission electron microscopy produce unusually anisotropic positron annihilation signals. We show that in experiments, the positron annihilation signals in $\beta$-Ga$_2$O$_3$ samples seem to be often dominated by split Ga vacancies.

\end{abstract}

\maketitle

\section{Introduction}

$\beta$-Ga$_2$O$_3$ is a direct wide bandgap (\SI{4.9}{\electronvolt}) semiconductor material with a high break-down electric field of \SI{8}{\mega\volt\per\centi\meter} and whose properties surpass those of GaN and SiC from the point of view of UV and high power applications \cite{pearton2018review}. As the already better characteristics are combined with the economical advantage of the availability of large size bulk crystals, \gao\ has earned a substantial research interest. The research efforts have led to controllable $n$-type conductivity via doping with Sn, Si or Ge \cite{pearton2018review} and unipolar field effect transistors and Schottky diodes have been developed \cite{devices2016}.

The rapid refinement of synthesis approaches of both epitaxial and bulk \gao~ has led to the situation where the crystalline quality of the material is no longer the main concern. Instead, recent reviews on \gao\ point out that the identification and control of dominating (point and extended) defects is the most important step for further improvement of the properties of \gao\ devices \cite{pearton2018review,mccluskey}. The low symmetry of the \gao\ structure is accompanied with anisotropic thermal, electrical and optical properties \cite{thermal_anisotropy2018, electrical_optical_anisotropy1997}. Theoretical work suggested already early on that the mono-vacancy defect of the lowest formation energy in \gao\ is of a peculiar type \cite{varley_splitvacancy}, where a cation atom neighbouring the cation mono-vacancy relaxes to an interstitial site half-way towards the vacancy \cite{varley_proton,lany2018}. These ``split'' Ga vacancies in which the open volume is split into two parts on either side of the interstitial, have been recently experimentally observed in infrared spectroscopy (hydrogenated form) and scanning transmission electron microscopy (STEM) studies \cite{weiser, stem}. 

Positron annihilation spectroscopy is a non-destructive method with selective sensitivity to neutral and negative vacancy-type defects, and second-order sensitivity to negatively charged defects without open volume \cite{tuomistomakkonen}. Thanks to these properties, positron annihilation methods have been successful in identifying the role of native point defects in the electrical compensation of $n$-type doped compound semiconductors such as GaN, ZnO, AlN, InN and In$_2$O$_3$ \cite{GaNPRL1997,ZnOPRL2003,MakiPRB2011,RauchPRB2011,KorhonenPRB2014}, as well as their alloys such as InGaN and AlGaN \cite{ChichibuNM2006,UedonoJAP2012,ProzheevaAPL2017,UedonoJAP2018,IshibashiJPCM2019,ProzheevaPRApplied2020}. In spite of the otherwise significant research interest in \gao, the number of reported studies with positron annihilation is surprisingly low \cite{ting_mseb_2002,korhonen2015,spie,islam2020}. These studies suggest that cation vacancies contribute to the compensation of $n$-type conductivity, but also point out a potential difficulty in interpreting the positron annihilation signals.

In this work, we present a comprehensive study of positron annihilation Doppler broadening signals in \gao, made possible by recent developments in theoretical calculation schemes. We show that the Doppler broadening signals emitted from positron-electron annihilations in \gao\ are characterized by an anisotropy of unprecedented magnitude for 3D crystals, and by relatively small differences between signals originating from the perfect lattice and various types of vacancy defects. Together, these effects prevent from employing the standard approaches of defect identification in semiconductors, where sample orientation does not need to be taken into account. By comparing to experiments, we show that the colossal signal anisotropy -- in particular the differences in the nature of the anisotropy -- contains information that can be used for defect identification even when a suitable reference material where the positron annihilation data could be interpreted as originating from the lattice only, typically referred to as "defect-free reference" is missing. We provide evidence of the experimental positron annihilation signals being in many cases dominated by the split Ga vacancies, supporting the recent findings \cite{weiser, stem, varley_splitvacancy, varley_proton}. It is likely that different levels of hydrogenation of these split Ga vacancies determine the level of electrical compensation in $n$-type \gao.

The remainder of this paper is organized as follows. Section II gives a brief account of the modeling scheme. In Section III, we calculate the positron annihilation signals in the \gao\ lattice and compare the observed anisotropy to that in other semiconductor crystals. Section IV presents the results obtained in 9 different vacancy defects, demonstrating the increase in anisotropy of the signals and full overlap with the lattice signals. In Section V we discuss the origins of the anisotropy in the electron-positron momentum density and the annihilation signals. In Section VI we compare the calculated signal anisotropies to those found in experiments, and find that the experimental signals are most likely dominated by the split Ga vacancy effects. We summarize our paper in Section VII. The Appendix provides the calculated ratio curves in all major lattice directions as well as ($S,W$) plots with alternative parametrization for the benefit of future defect identification endeavours in \gao.

\section{Modeling positron states and annihilation in {$\boldsymbol{\beta}$-G\lowercase{a}$_2$O$_3$}}

We study the $\beta$-Ga$_2$O$_3$ lattice and its defects using density-functional calculations and the Heyd-Scuseria-Ernzerhof (HSE06) screened hybrid functional with a modified fraction of Hartree-Fock exchange of 35\% consistent with earlier work~\cite{varley_splitvacancy}, and include the semi-core Ga $3d$ electrons as explicit valence states. We use monoclinic 160-atom supercells to describe the previously considered defect models~\cite{varley_splitvacancy,varley_proton}. We apply the \textsc{vasp} code~\cite{KressePRB1996,KresseCMS1996,KressePRB1999} and its implementation of the projector augmented-wave (PAW) method \cite{BlochlPRB1994}. Tests made using a simpler and more affordable model, the generalized gradient approximation (GGA) by Perdew, Burke and Ernzerhof (PBE) \cite{PBE} indicated that HSE06 and PBE give very consistent results for the same ionic structure. We used HSE06 in order to be consistent with earlier work and the defect structures used~\cite{varley_splitvacancy,varley_proton}. In future work we might have to consider larger defect structures using larger supercells but then we can safely resort to PBE.

We model delocalized and trapped positrons and their lifetimes using the two-component density functional theory for electron-positron systems and approximating the correlation potential and enhancement factor using the local density approximation (LDA) ~\cite{BoronskiPRB1986}. We assume that the localized positron does not affect the average electron density and apply the zero-positron-density limits of the functionals. This method, which is justified for delocalized positrons, has been shown to give also for localized positrons results that are in agreement with more self-consistent modeling~\cite{PuskaPRB1995}. The LDA enhancement typically predicts too high annihilation rates with 3d electrons in comparison with the experiment, but differences (or ratios) in lifetimes and Doppler spectra can be compared with high confidence \cite{tuomistomakkonen,MakkonenJPCS2005,MakkonenPRB2006}.

In order to confirm a bound positron state and assess the role of the finite size of the supercells used, we have analyzed the localization of the positron density as well as the dispersion of the positron band. The borderline cases, in which the 160-atom cell alone does not provide conclusive results, include the hydrogenated vacancies $V_{\text{Ga}}^{\text{ib}}$-2H and $V_{\text{Ga}}^{\text{ic}}$-2H, in which the H atoms reduce the open volumes at the defect site. For these, we have made checks using the PBE GGA functional and supercells up to 960 atoms. For the standard 160-atom supercells we apply an extrapolation technique based on sampling the positron band at 2 $\mathbf{k}$ points in the case of defects localizing the positron \cite{KorhonenPRB1996}.

\begin{figure}[h]
  \centering
    \includegraphics[width=\linewidth]{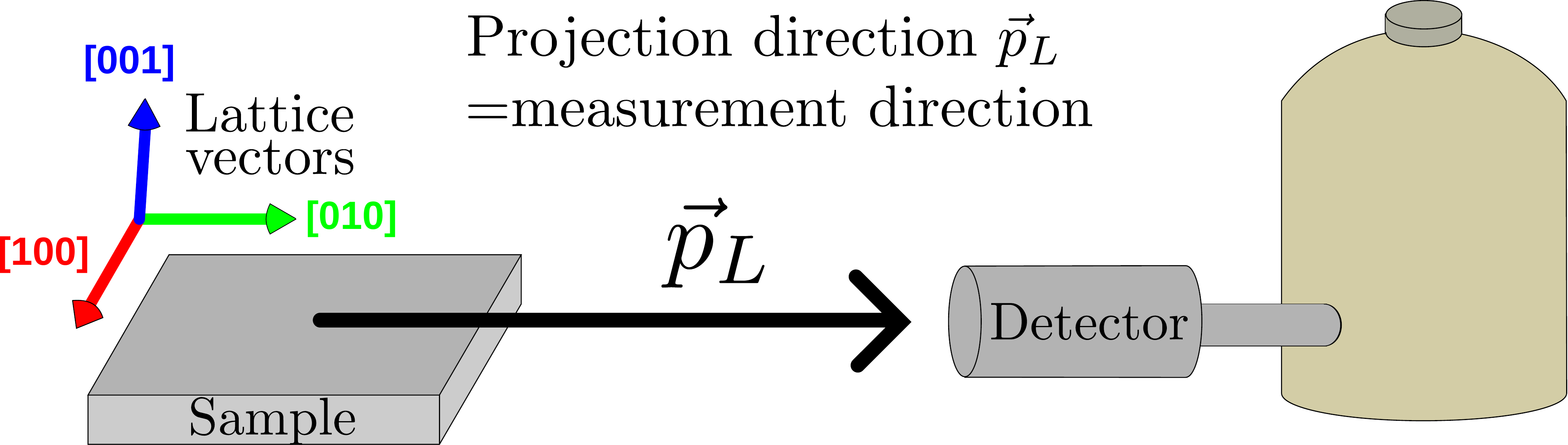}
    \caption{Illustration of the projection direction $\mathbf{p}_L$ with respect to the experiment geometry.}
     \label{f:directions}
     \end{figure}

We model the momentum densities of annihilating electron-positron pairs using the model by Alatalo and coworkers~\cite{AlataloPRB1996}, a $2\times 2\times 2$ $\Gamma$-centered mesh and reconstructed PAW wave functions~\cite{MakkonenJPCS2005,*MakkonenPRB2006}. In order to be able to compare with Doppler broadening experiments, the  3-dimensional (3D) momentum density $\rho (\mathbf{p})$ is first projected into the desired crystal direction (as illustrated in Fig. \ref{f:directions})
to give the Doppler spectrum,
\begin{align}\label{projeq}
    \rho (p_{L})=\int\int \rho (\mathbf{p}) dp_{x}dp_{y},
\end{align}
which is then convoluted with the experimental resolution function (a Gaussian with a full width at half maximum corresponding to 0.95 keV or 1.25 keV, in case of two-detector coincidence and one-detector measurements used to obtain the shape parameters ($S,W$), respectively.) In order to be able to consider the monoclinic cell and 
projections to any lattice direction (arbitrary $\mathbf{p}_{L}$), 
we have implemented in our code the tetrahedron projection algorithm of Matsumoto, Tokii and Wakoh~\cite{MatsumotoJPSJ2004}.  

The anisotropy observed in Doppler spectra $\rho (p_{L})$ as a function of the longitudinal direction $\mathbf{p}_{L}$ is a consequence of the anisotropy of the 3D momentum density of annihilating electron-positron pairs, $\rho (\mathbf{p})$ (see Eq.~(\ref{projeq})). For a wide band gap material, it is the shape of the first Brillouin zone and the  intensities of the higher Umklapp momentum components that determine the anisotropy of the Doppler spectra. In the case of a positron delocalized in a defect-free lattice, there exists a relatively close analogy to the 3D electron momentum density and its projections, the Compton profiles. The positron modifies these quantities through its own momentum density and by emphasizing the role of valence electron orbitals, with which it is mostly annihilating. Vacancies trapping positrons, however, are different as the positron selectively samples only the orbitals involving the neighboring ions. $\beta$-Ga$_2$O$_3$ is characterized by a large anisotropy for both delocalized and localized positrons, which is highly unusual. How exactly the anisotropy of the 3D momentum density of annihilating electron-positron pairs arises from the local ionic and electronic structures of the defects, is not yet fully understood in the case of $\beta$-Ga$_2$O$_3$. For this reason and to make a closer connection with experiment, we focus on the Doppler spectra. The detailed analysis of the microscopic origin of the anisotropy for vacancy defects could in principle be possible in future studies by enabling visualization of site and angular momentum dependent decompositions of the 3D momentum density of annihilating pairs.

\section{Positrons in the $\boldsymbol{\beta}$-G\lowercase{a}$_2$O$_3$ lattice}
\label{lattice}

\begin{figure}[h]
  \centering
  \begin{subfigure}[b]{0.49\linewidth}
    \includegraphics[width=\linewidth]{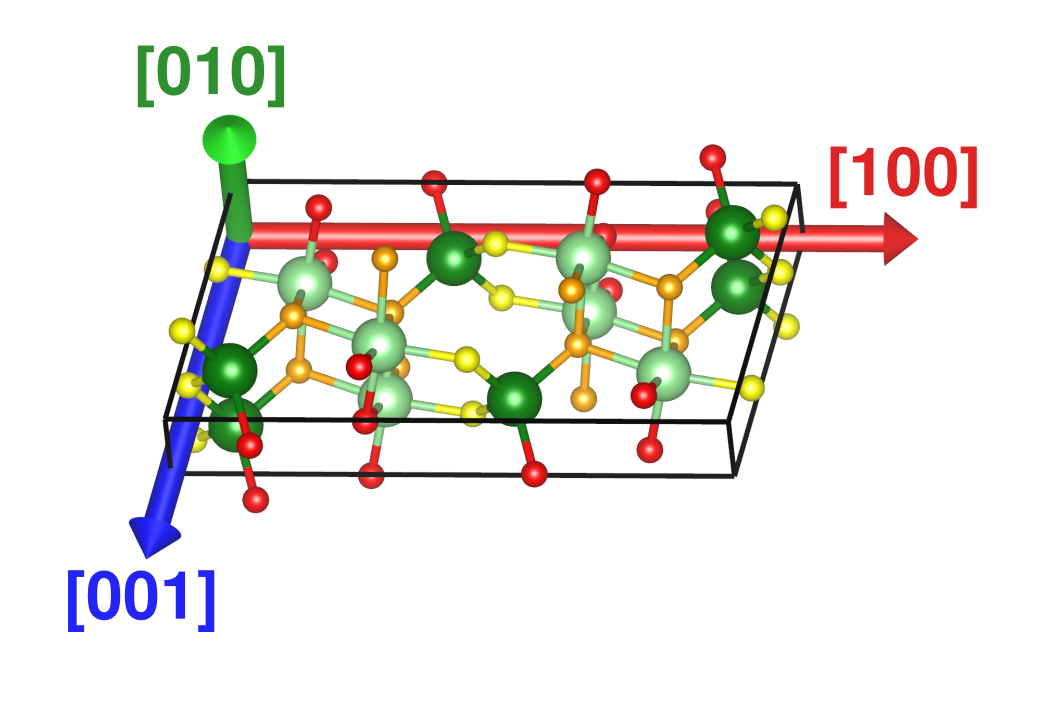}
    \caption{$\beta$-Ga$_2$O$_3$ unit cell }
     \label{f:unit_cell}
     \vspace{10pt}
  \end{subfigure}
    \centering
    \begin{subfigure}[b]{0.49\linewidth}
        \includegraphics[width=\linewidth]{ga2o3_paper_010.pdf}
        \caption{cross section perpendicular to [010]}
        \label{f:crosssection_010}
    \end{subfigure}

        \centering
    \begin{subfigure}[b]{0.35\linewidth}
        \includegraphics[width=\linewidth]{ga2o3_paper_100.pdf}
        \caption{cross section perpendicular to [100]}
        \label{f:crosssection_100}
    \end{subfigure}
~
\begin{subfigure}[b]{0.35\linewidth}
       \includegraphics[width=\linewidth]{ga2o3_paper_001.pdf}
        \caption{cross section perpendicular to [001]}
        \label{f:crosssection_001}
    \end{subfigure}
~
\begin{subfigure}[b]{0.2\linewidth}

        \includegraphics[width=\linewidth]{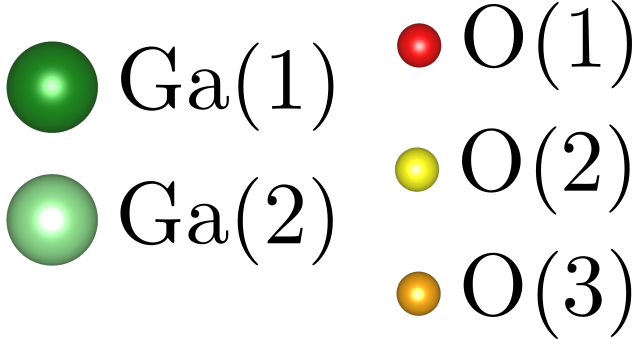}
\vspace{1.2cm}

    \end{subfigure}
        
        \caption{The standard conventional unit cell of $\beta$-Ga$_2$O$_3$ and cross sections perpendicular to lattice vectors \cite{vesta}.}
        \label{f:structure}
\end{figure}

$\beta$-Ga$_2$O$_3$ is the most stable phase of Ga$_2$O$_3$ and it has a base-centered monoclinic crystal structure (space group C2/m) with relatively low symmetry. The primitive unit cell of $\beta$-Ga$_2$O$_3$ consists of 10 atoms but due to its difficult shape the \textit{standard conventional} unit cell is typically used. The standard conventional unit cell is almost rectangular apart from one angle of \SI{103.7}{\degree} (between the [100] and [001] lattice vectors) and consists of 20 atoms (Fig. \ref{f:unit_cell}). We use the standard conventional unit cell throughout this paper, and refer to it as the \textit{unit cell}. Due to the non-orthogonal structure, the lattice planes described by Miller indices (100) and (001) are not parallel to the planes spanned by the lattice vectors and, for the sake of simplicity, we define lattice planes by the two lattice vectors spanning the plane.

The low symmetry of the lattice is well visible in the atomic structure as the lattice appears clearly different from all three lattice vector directions (see Fig. \ref{f:structure}). Note that the structure has open ``channels'' in the direction of the [010] lattice vector while the cross sections perpendicular to the [100] and [001] lattice vectors appear clearly denser. There are two inequivalent Ga sites in $\beta$-Ga$_2$O$_3$ with 4 and 6 nearest neighbour oxygen atoms (Ga(1) and Ga(2), respectively) and three inequivalent oxygen sites. One of the rare symmetries \gao~ lattice has is a \SI{180}{\degree} rotation with the [010] lattice direction as the rotation axis. 
Following this symmetry, the unit cell has all these five inequivalent sites (two Ga and three O) in two different orientations twice.

\begin{figure}[tb]
  \centering
  \begin{subfigure}[b]{0.49\linewidth}
    \includegraphics[width=\linewidth]{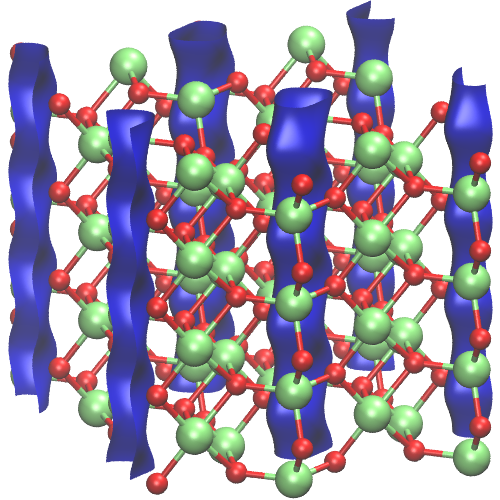}
    \caption{$\beta$-Ga$_2$O$_3$}
     \label{f:ga2o3_latticestate}
  \end{subfigure}
    \centering
  \begin{subfigure}[b]{0.49\linewidth}
    \includegraphics[width=\linewidth]{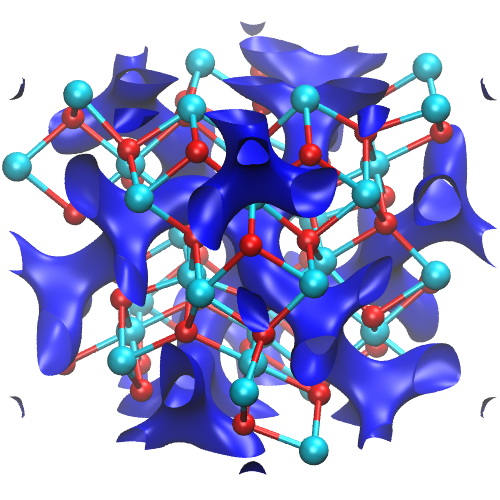}
    \caption{In$_2$O$_3$}
     \label{f:in2o3_latticestate}
  \end{subfigure}

    \begin{subfigure}[b]{0.49\linewidth}
        \includegraphics[width=\linewidth]{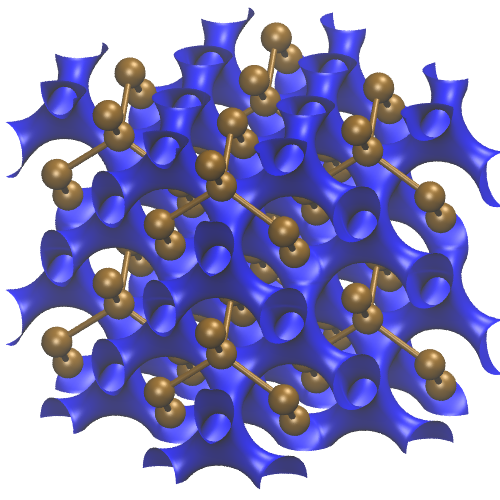}
        \caption{Si}
        \label{f:si_latticestate}
    \end{subfigure}
    \centering
    \begin{subfigure}[b]{0.49\linewidth}
        \includegraphics[width=\linewidth]{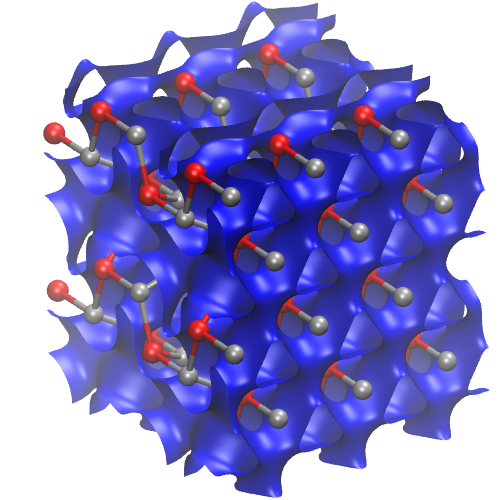}
        \caption{ZnO}
        \label{f:zno_latticestate}
    \end{subfigure}
        \caption{Positron lattice state densities in $\beta$-Ga$_2$O$_3$, In$_2$O$_3$, Si and ZnO (positron density in blue, oxygen atoms in red) \cite{vmd, tachyon}. The positron density isosurface values are chosen in a way that best illustrates the general behavior of the positron density (Ga2O3: 63\%, In2O3: 73\%, Si: 79\%, ZnO: 50\%). Typically, the delocalized positron lattice state is three-dimensional but in $\beta$-Ga$_2$O$_3$ the positron density forms "tubes" along the [010] lattice vector.}
        \label{f:latticestates}
\end{figure}

The positron density of the delocalized state in the $\beta$-Ga$_2$O$_3$ lattice is shown in Fig. \ref{f:latticestates} together with In$_2$O$_3$, Si and ZnO for comparison. Interestingly, the positron density in the $\beta$-Ga$_2$O$_3$ lattice forms one-dimensional tubes along the [010] lattice vector, while in the other 3D crystalline structures in Fig. \ref{f:latticestates} the positron density forms three-dimensional networks. A similar (but two-dimensional) delocalized positron state is known to exist in layered lattice structures such as graphite \cite{graphite, graphite2,TangPRB2002}, while one-dimensional positron states have been proposed to exist in carbon nanotubes \cite{nanotubes}.

\begin{figure}[h]
\begin{center}
\includegraphics{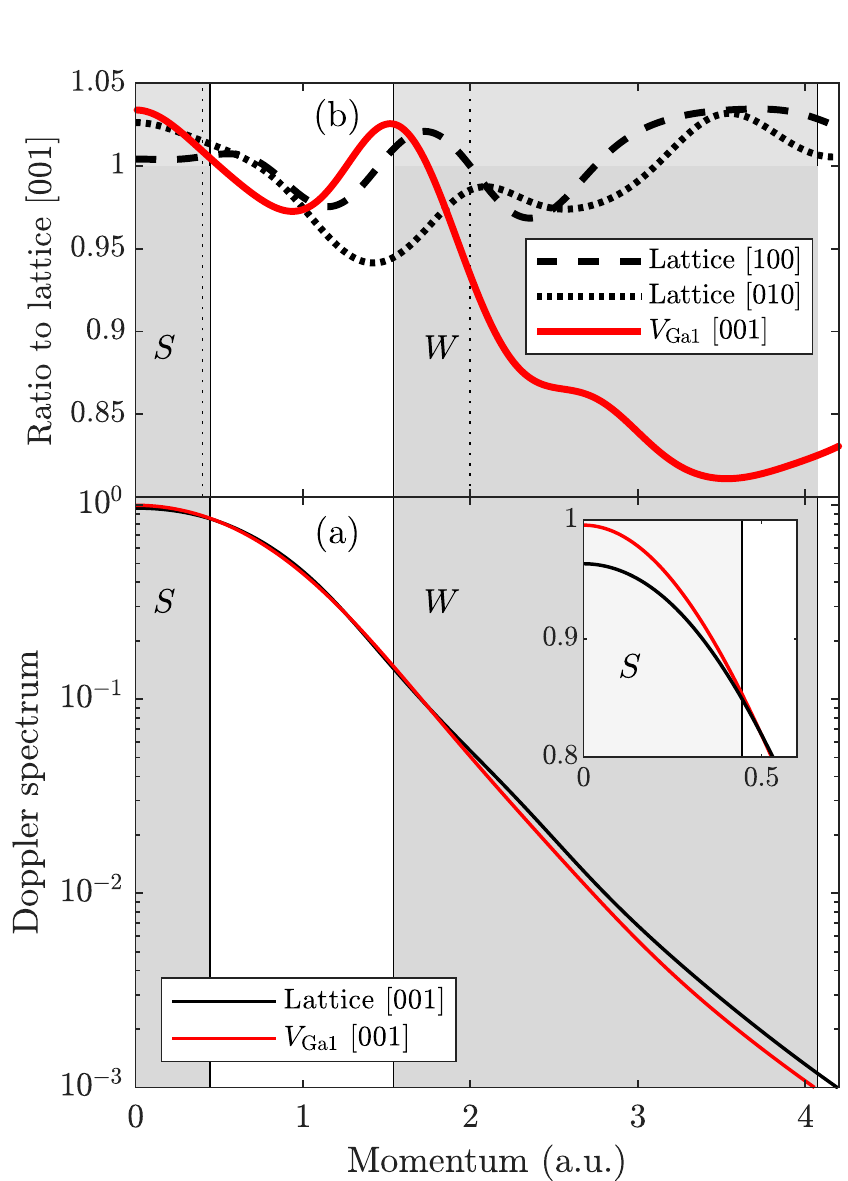}    
\end{center}
\caption{(a) Calculated Doppler spectra in the $\beta$-Ga$_2$O$_3$ lattice and $V_{\mathrm{Ga1}}$ in the [001] lattice direction. The inset shows a magnification of the $S$ parameter region on the linear scale. (b) Ratios of the calculated Doppler spectra in the $\beta$-Ga$_2$O$_3$ lattice in the [100] and [010] lattice directions, and $V_{\mathrm{Ga1}}$ in the [001] lattice direction, normalized by the lattice Doppler spectrum in the [001] lattice direction. The shaded regions show the integration windows of the ($S,W$) parameters, and the dotted vertical lines represent alternative ($S,W$) parameter windows.}
\label{f:bulk_tiled}
\end{figure}

Fig. \ref{f:bulk_tiled}a shows the calculated momentum distribution of annihilated electron-positron pairs (Doppler broadened spectrum) in the [001] lattice direction of the $\beta$-Ga$_2$O$_3$ lattice. The calculation for the missing Ga(1) atom (Ga(1) vacancy), denoted by $V_{\text{Ga1}}$, is shown for comparison. Changes in the Doppler spectrum are best monitored by so-called ratio curves where the Doppler spectra are normalized by a common reference spectrum due to the differences being on the percent-level in a signal whose intensity covers several orders of magnitude. Fig. \ref{f:bulk_tiled}b shows the calculated lattice Doppler spectra in the [100] and [010] directions as well as the $V_{\text{Ga1}}$ spectrum in the [001] direction, all normalized by the lattice spectrum in the [001] direction. The potential difficulty in distinguishing between the vacancy and a different lattice orientation is evident, as the lattice spectrum in the [010] direction has very similar intensity in the low momentum range (the $S$ parameter region) as the $V_{\text{Ga1}}$ spectrum in the [001] direction. In addition, the high-momentum range, even if clearly different for these two when presented as in Fig.~\ref{f:bulk_tiled}a, becomes very similar upon integration due the rapidly decreasing signal intensity in the $W$ parameter region. A detailed experimental identification would require having at hand a well-specified and well-characterized experimental reference that -- for the time being -- is not the case for \gao. The differences between the lattice spectra in [001] and [100] directions are clearly smaller.

Instead of analyzing the full ratio curves, the shape of the Doppler spectrum is often described with integrated shape parameters ($S,W$) due to experimental (time) constraints. These parameters are defined as the fraction of annihilated electron-positron pairs in low ($S$) and high ($W$) momentum regions, shown as the shaded areas in Fig.~\ref{f:bulk_tiled}. The intensity of a Doppler spectrum decreases rapidly towards higher momenta (as seen in Fig. \ref{f:bulk_tiled}a), and especially in the $W$ parameter region the first half atomic units (a.u.) of the $W$ window contain approximately half of the signal weight (number of counts in the experiments) of the $W$ parameter. Note that in the ratio curves 
(Fig. \ref{f:bulk_tiled}b) the signal intensity is normalized to 1 and the (relative) ($S,W$) parameters cannot be determined by directly integrating the data in the ($S,W$) windows of a ratio plot.

The $S$ parameter window is typically set in such a way that approximately half of the distribution weight (half of the counts in the experiment) is within the window (Fig. \ref{f:bulk_tiled}), to retain the statistical accuracy of collecting a large number of counts in the experiment. The lower limit of the $W$ parameter integration window is chosen far enough from the peak center in order to have a minimal contribution from the "free-electron" distribution that dominates the $S$ parameter region (see Fig.~7 of Ref. \onlinecite{tuomistomakkonen}). However, pushing the lower limit of the $W$ parameter too far quickly deteriorates the statistical accuracy of the parameter due to the close-to-exponentially decreasing signal intensity in this range. The ($S,W$) parameter windows used in this work are shown in the figure with shaded areas: the $S$ parameter window ranges from 0 to 0.45 atomic units (a.u., corresponding to $0 - 0.83$~keV) and the $W$ window from 1.54 to 4.07 a.u. ($2.87- 7.60$~keV). The dotted lines in Fig.~\ref{f:bulk_tiled}b represent narrower ($S,W$) windows, discussed in the Appendix. We wish to stress that while the ($S,W$) windows should be optimized for each material \cite{LinezPRB2016}, in practice they often are not and "standard" windows are used instead, and the ($S,W$) windows used in this work are similar to the "standard" windows. At this point we also wish to point out the fact that the ($S,W$) parameters only describe the shape of the Doppler spectrum, without any direct physical interpretation.

\begin{figure}[htbp]
  \centering
  \begin{subfigure}[b]{\linewidth}
\includegraphics[height=0.23\textheight]{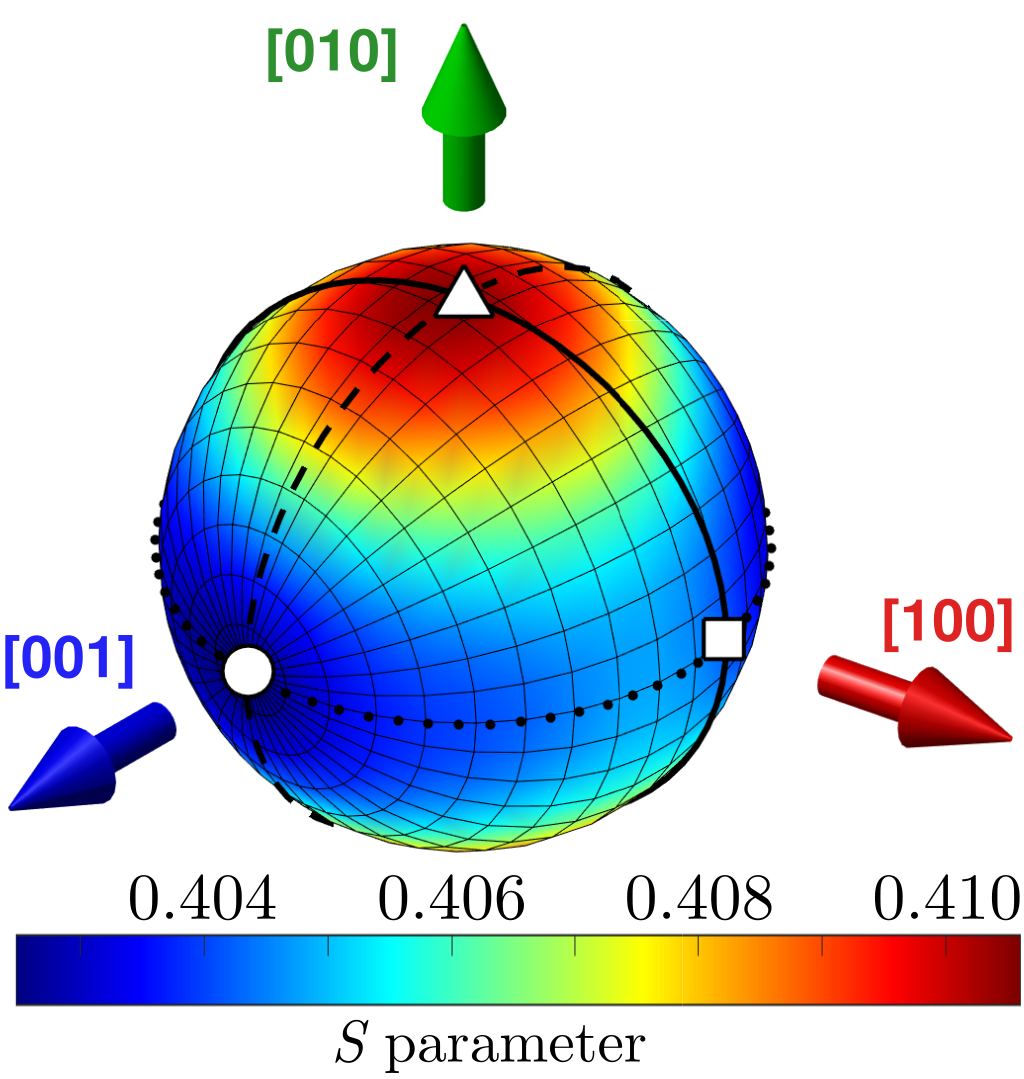}
        \caption{Full anisotropy of the $S$ parameter with respect to the $\beta$-Ga$_2$O$_3$ lattice vectors.}
    \label{f:bulk_s_sphere}
  \end{subfigure}

  \begin{subfigure}[b]{\linewidth}
\includegraphics[height=0.23\textheight]{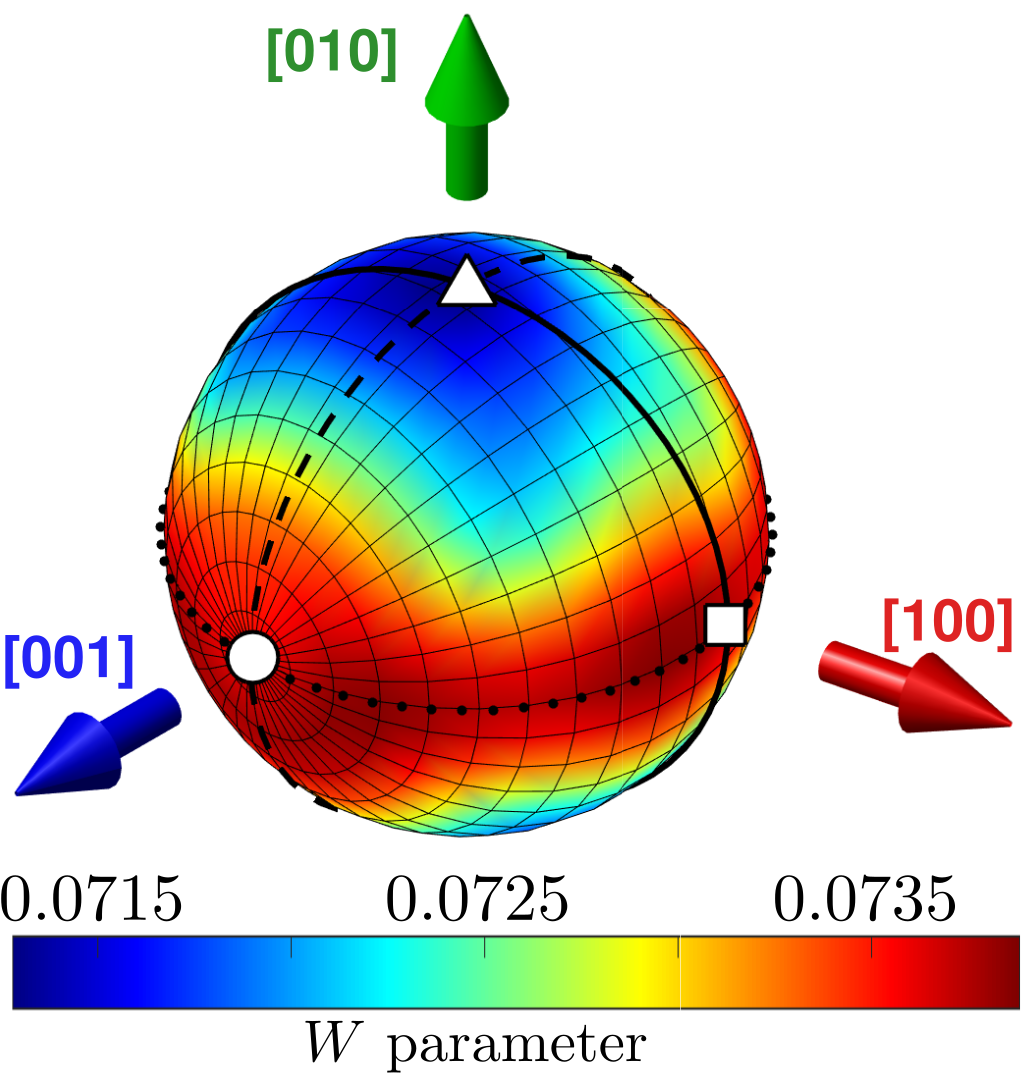}
        \caption{Full anisotropy of the $W$ parameter with respect to the $\beta$-Ga$_2$O$_3$ lattice vectors.}
    \label{f:bulk_w_sphere}
  \end{subfigure}

\centering
  \begin{subfigure}[b]{\linewidth}
    \includegraphics{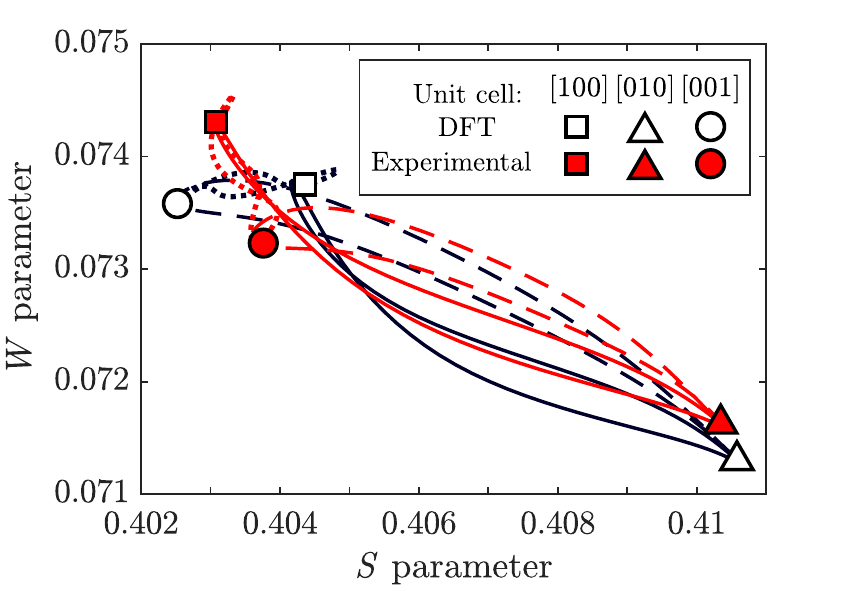}
    \caption{The Doppler parameters $(S,W)$ calculated in the \gao~ lattice using two different unit cells, a computational \cite{varley_splitvacancy} and experimental unit cell \cite{ahman1996}. The slightly different atomic locations lead to differences in calculations in the [001] and [100] directions.}

     \label{f:bulk_sw}
  \end{subfigure}
  \caption{Anisotropy of the Doppler signal of the $\beta$-Ga$_2$O$_3$ lattice. The full, dashed and dotted curves along the geodesics between the lattice directions [001], [010] and [100] are the same in (a), (b) and (c). The coloration of the lattice direction arrows follows the convention of Fig.~\ref{f:structure}.}
\label{f:bulk_anisotropy}     
  \end{figure}

The full three-dimensional anisotropies of the $S$ and $W$ parameters in the $\beta$-Ga$_2$O$_3$ lattice are shown in Figs.~\ref{f:bulk_s_sphere} and \ref{f:bulk_w_sphere}. The highest $S$ (lowest $W$) parameter lies in the [010] direction ($S=0.411$) and it decreases ($W$ increases) towards the plane spanned by the [100] and [001] lattice vectors. At angles less than \SI{45}{\degree} from the [010] lattice vector, the $S$ parameter does not change significantly with rotation around the [010] lattice vector. At angles higher than \SI{45}{\degree} from the [010] lattice vector the $S$ parameter starts to visually differ from the apparent rotational symmetry and, in the [100]-[001] plane (\SI{90}{\degree} from [010]) the $S$ parameter is the highest in the vicinity of the [100] ($S=0.405$) direction and the smallest in the direction of the [001] lattice vector ($S=0.403$). The $W$ parameter behaves in a roughly opposite way to the $S$ parameter, but with less rotational symmetry around the [010] lattice vector and, in the [100]-[001] plane, the $W$ parameter is constant. While Figs.~\ref{f:bulk_s_sphere} and \ref{f:bulk_w_sphere} may be visually appealing, comparing the full three-dimensional anisotropies in this way is not optimal. It turns out that the interesting features of the ($S,W$) parameters -- such as maxima and minima -- tend to be located either at the lattice vector directions or on the geodesics connecting them. Hence, we limit our discussion to these directions, as shown in Figs.~\ref{f:bulk_s_sphere}, \ref{f:bulk_w_sphere} and~\ref{f:bulk_sw}. The notation of Fig. \ref{f:bulk_sw} is used throughout the discussion: the lattice directions [100], [010] and [001] are represented by square, triangle and circle, and the geodesics between [100] and [010] are described with a full curve, between [010] and [001] with a dashed curve, and between [001] and [100] with dotted curve, as they appear on the sphere in Figs.~\ref{f:bulk_s_sphere} and \ref{f:bulk_w_sphere}. In Fig. \ref{f:bulk_sw}, the full, dashed and dotted curves make loops which intersect two out of the three lattice directions, as do the geodesics. The lack of mirror symmetries in the \gao\ lattice (see Fig.~\ref{f:structure}) is at the origin of the non-coinciding ($S,W$) parameters when going from, e.g., [100] to [010] and then from [010] to [-100]. The loops return to their initial ($S,W$) already after a \SI{180}{\degree} rotation (in real space) as a result of the intrinsic inversion symmetry of Doppler broadening. 

Figure \ref{f:bulk_sw} shows the ($S,W$) parameters in the $\beta$-Ga$_2$O$_3$ lattice. For completeness, we show results calculated using two different published \gao\ unit cells: the computationally determined unit cell \cite{varley_splitvacancy} used throughout this work and an experimental unit cell determined in Ref.~\onlinecite{ahman1996}. The exact locations of atoms in these two unit cells are slightly different and, unlike in materials with cubic or hexagonal crystal structures, ideal atomic structures based on symmetry cannot be used in the modeling. Clearly detailed X-ray experiments are needed on state-of-the-art \gao\ materials to resolve the atomic-level details of the crystal structure, which may not be the same for, \textit{e.g.}, strained thin films and single crystals. The differences in the calculated positron signals using the two cells, in particular in the [100] and [001] directions, manifests the sensitivity of the Doppler signal anisotropy in these lattice directions to the exact atomic structure. This is likely to lead to certain challenges in the comparison between theory and experiments. For consistency, we use the computationally determined unit cell in all the presented modeling data in this work.

The ($S,W$) parameters in the [100] and [001] lattice directions are very similar (DFT unit cell data in Fig~\ref{f:bulk_sw}), the $S$ parameters differ by a factor of 1.005 and the $W$ parameters are practically identical, while the [010] direction has much higher $S$ and smaller $W$ parameters. The extrema of the Doppler signal anisotropy are approximately at the lattice vector directions. Along the geodesics between the lattice vectors, ($S,W$) deviate somewhat from linear interpolation. The range of the Doppler signal anisotropy in the $\beta$-Ga$_2$O$_3$ lattice is 1.000-1.020 in the $S$ parameter and 0.97-1.00 in the $W$ parameter for the selected integration windows. We define the anisotropy range as the maximum and minimum $S$ ($W$) parameter divided by the $S$ ($W$) of the $\beta$-Ga$_2$O$_3$ lattice in the [001] lattice direction (the direction of the smallest calculated $S$ parameter in the $\beta$-Ga$_2$O$_3$ lattice). We use these values as the reference ($S,W$) point throughout the discussion.

\begin{figure}[h]
  \begin{center}
\includegraphics{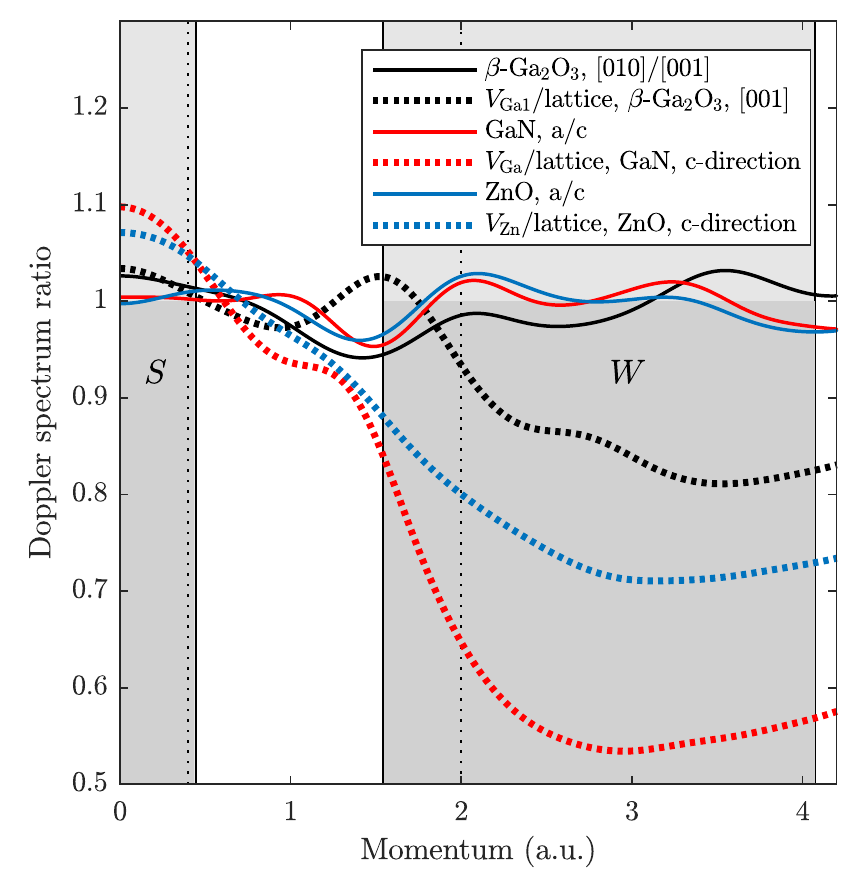}
\caption{Doppler signal ratios for the lattice between different projection directions as well as monovacancy-lattice ratios in $\beta$-Ga$_2$O$_3$, GaN and ZnO.}
\label{f:material_ratios}
\end{center}
\end{figure}

\begin{table}[h]
  \caption{Comparison of Doppler signal anisotropies in lattice and monovacancy-lattice ratios in $\beta$-Ga$_2$O$_3$, GaN and ZnO.}
  \label{t:material_anisotropies}
\begin{center}
  \begin{tabular}{lcccc}
\hline
\hline
& \multicolumn{2}{c}{Lattice anisotropy}  & \multicolumn{2}{c}{Vacancy/lattice-ratio} \\
\\
    & $S_\mathrm{[010]}$/$S_\mathrm{[001]}$ & W$_\mathrm{[010]}$/$W_\mathrm{[001]}$& $S^\mathrm{vac}_\mathrm{[001]}$/$S_\mathrm{[001]}$&  $W^\mathrm{vac}_\mathrm{[001]}$/$W_\mathrm{[001]}$\\
    \hline
    Ga$_2$O$_3$ & 1.020  & 0.97 & 1.020 & 0.96  \\
\multicolumn{1}{c}{}\\
    & $S_\mathrm{a}$/$S_\mathrm{c}$ & W$_\mathrm{a}$/$W_\mathrm{c}$& $S^\mathrm{vac}_\mathrm{c}$/$S_\mathrm{c}$&  $W^\mathrm{vac}_\mathrm{c}$/$W_\mathrm{c}$\\
    \hline
    GaN & 1.003 & 0.99 & 1.071 &	0.71 \\
ZnO &  1.004 & 1.00 & 1.057 &	0.82 \\
\hline
\hline
\end{tabular}
\end{center}
\end{table}

It is worth noting that the Doppler broadening signals from crystal lattices are anisotropic by nature and observed in, e.g., Si \cite{Si_anisotropy} and ZnO \cite{ZnO_anisotropy}, and that it is the magnitude of the phenomenon that is unusual in \gao. Figure \ref{f:material_ratios} shows the calculated Doppler ratio curve anisotropies and the monovacancy/lattice ratios for wurtzite GaN, wurtzite ZnO and \gao. It is clearly seen that the $S$ parameter region hardly changes when examined along the \textit{a} or \textit{c} directions of the wurtzite lattice of GaN or ZnO, and, while the differences are larger in the $W$ parameter region, the ratio is still very close to unity. In contrast, the monovacancy/lattice ratios are very different from the respective lattice anisotropies both for $V_{\text{Zn}}$ in ZnO and for $V_{\text{Ga}}$ in GaN.

Table \ref{t:material_anisotropies} shows the changes in the values of the ($S,W$) parameters for the lattice anisotropies and the monovacancy/lattice comparisons. In GaN and ZnO, the \textit{a}/\textit{c}-ratios of the lattice Doppler signal are less than 1.005 in the $S$ and 0.99 in the $W$ parameter, and the anisotropies (not shown) of the monovacancy signals in GaN and ZnO are similar or smaller. The differences between monovacancy and lattice are more than tenfold compared to the anisotropy in these two materials: the calculated monovacancy/lattice ratio in GaN is 1.071 in the $S$ and 0.71 in the $W$ parameter, and in ZnO they are 1.057 and 0.82, respectively. If the positron signal anisotropy is significantly smaller than the vacancy/lattice signal ratio, the anisotropy does not affect the interpretation of the experimental results in practice and can be mostly disregarded in the analysis. However, in $\beta$-Ga$_2$O$_3$, the monovacancy-lattice signal ratio is much smaller than that in GaN and ZnO, while the positron signal anisotropy in \gao\ is an order of magnitude stronger than in GaN or ZnO. Even in Si, where the difference between the [100] and [110] lattice directions is larger than in the wurtzite compounds \cite{Si_anisotropy}, the anisotropy is vanishingly small compared to that found in \gao. As a result, the Doppler signal anisotropy is of the same magnitude as the monovacancy/lattice-ratio implying that \gao\ needs to be treated differently than other semiconductor materials.

\section{Positrons in vacancy defects of  $\boldsymbol{\beta}$-G\lowercase{a}$_2$O$_3$}

\subsection{Structure of vacancies}

\begin{figure}[htb]
  \centering
  \includegraphics[width=\linewidth]{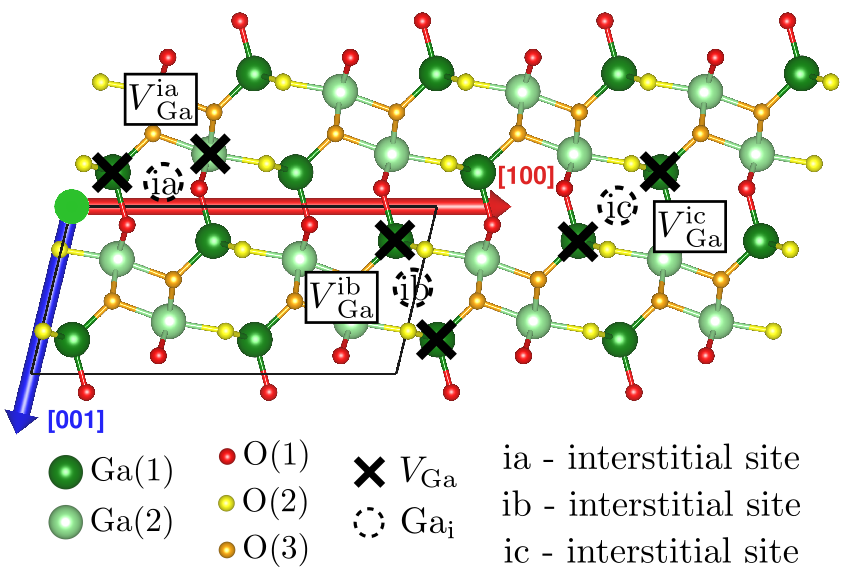}
  \caption{The structure of split-vacancies in $\beta$-Ga$_2$O$_3$ \cite{vesta}.}
  \label{f:vacancy_structure}
\end{figure}

In spite of $\beta$-Ga$_2$O$_3$ consisting of only two different elements, it hosts a wide variety of different monovacancy-size defects due to the inequivalent Ga and O sites. As the oxygen monovacancies were found not to trap positrons in the calculations as is usual for oxides due the small size of the open volume \cite{MakkonenJPCMoxides}, we focus on cation monovacancies in the following. The two inequivalent Ga sites, the four-fold coordinated Ga(1) and the six-fold coordinated Ga(2) of the \gao\ unit cell immediately lead to two different Ga monovacancies $V_{\mathrm{Ga1}}$ and $V_{\mathrm{Ga2}}$. The cation monovacancies in $\beta$-Ga$_2$O$_3$ have a special property that the regular monovacancies, $V_{\mathrm{Ga1}}$ and $V_{\mathrm{Ga2}}$, can relax into three different configurations $V_{\mathrm{Ga}}^{\mathrm{ia}}$, $V_{\mathrm{Ga}}^{\mathrm{ib}}$ and $V_{\mathrm{Ga}}^{\mathrm{ic}}$ \cite{varley_splitvacancy, varley_proton}. In the relaxation process, a neighbouring four-fold coordinated Ga(1) atom relaxes inwards into the interstitial space. The resulting split Ga vacancy has an open volume on both sides of the center interstitial, resulting in two "half-vacancies". The split Ga vacancy $V_{\mathrm{Ga}}^{\mathrm{ia}}$ forms at Ga(1) and Ga(2) sites, while $V_{\mathrm{Ga}}^{\mathrm{ib}}$ and $V_{\mathrm{Ga}}^{\mathrm{ic}}$ are formed by two Ga(1) sites, as illustrated in Fig. \ref{f:vacancy_structure}. The split Ga vacancies classify as mono-vacancies as they consist of only one missing atom.

\begin{figure}[htb]
  \centering
  \includegraphics[width=0.79\linewidth]{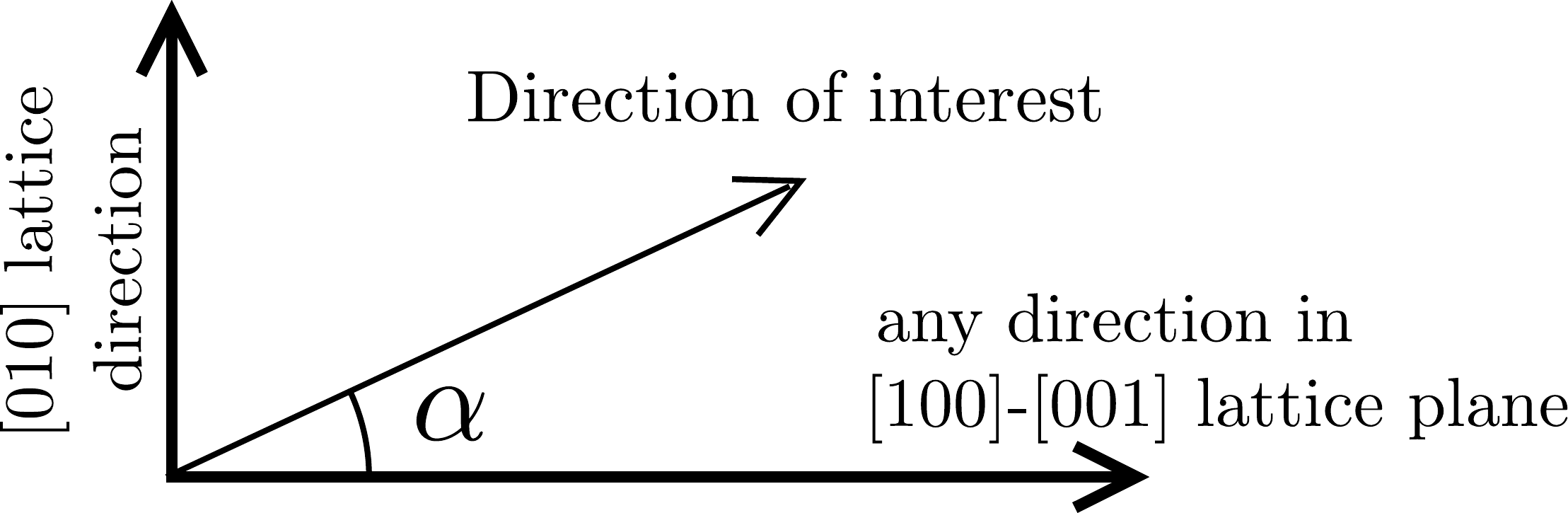}
  \caption{Illustration of the angle $\alpha$ in Eq. \ref{e:correction} for taking account the Doppler signals of defects in two orientations.}
  \label{f:alpha}
\end{figure}

Due to the symmetry of the \gao~lattice with respect to the \SI{180}{\degree} rotation with [010] as the rotation axis, all vacancy structures in \gao~ appear in these two orientations. 
The Doppler broadening signal is intrinsically inversion symmetric, rendering the Doppler signals of both orientations identical in the [010] lattice direction and the [100]-[001] lattice plane, but does not cancel all the differences in the directions between the [100]-[001] lattice plane and the [010] lattice direction. The presence of defects in these two orientations can be taken into account by calculating the Doppler broadening signal for one of these orientations and applying a simple correction based on symmetry. Assuming that defects appear with equal probability in both of these identical orientations, a correction for the Doppler signal, $DB$, can be calculated for any given direction with an angle $\alpha$ from [100]-[001] lattice plane (see Fig. \ref{f:alpha}). The corrected signal $DB_{\mathrm{corr}}$ is a mirror average with respect to the [010] lattice vector:
\begin{equation}
  DB_{\mathrm{corr}}(\alpha)=\frac{DB_{\text{uncorr}}(\alpha) + DB_{\text{uncorr}}(\pi-\alpha)}{2}, \label{e:correction}
\end{equation}
where $DB_{\mathrm{uncorr}}$ stands for the uncorrected calculated signal.

\subsection{Positron states}

We calculated the positron states in 9 cation vacancy systems that can be divided into three groups: regular mono-vacancies $V_{\mathrm{Ga1}}$ and $V_{\mathrm{Ga2}}$, split Ga vacancies $V_{\mathrm{Ga}}^{\mathrm{ia}}$, $V_{\mathrm{Ga}}^{\mathrm{ib}}$ and $V_{\mathrm{Ga}}^{\mathrm{ic}}$, and hydrogenated split Ga vacancies $V_{\mathrm{Ga}}^{\mathrm{ib}}$-1H, $V_{\mathrm{Ga}}^{\mathrm{ib}}$-2H, $V_{\mathrm{Ga}}^{\mathrm{ic}}$-1H and $V_{\mathrm{Ga}}^{\mathrm{ic}}$-2H. We chose systems that are simple and predicted to be energetically favorable ($V_{\mathrm{Ga}}^{\mathrm{ia}}$, $V_{\mathrm{Ga}}^{\mathrm{ib}}$, $V_{\mathrm{Ga}}^{\mathrm{ic}}$) \cite{varley_proton}, or suggested via experiments ($V_{\mathrm{Ga}}^{\mathrm{ib}}$, $V_{\mathrm{Ga}}^{\mathrm{ic}}$, $V_{\mathrm{Ga}}^{\mathrm{ib}}$-2H) \cite{weiser, stem}. 
At Fermi level close to the conduction band, the charge state of "clean" cation mono-vacancies is predicted as $-3$, after passivation with one hydrogen ($V_{\mathrm{Ga}}^{\mathrm{ib}}$-1H and $V_{\mathrm{Ga}}^{\mathrm{ic}}$-1H) the charge state is predicted as $-2$, and with 2 hydrogen atoms ($V_{\mathrm{Ga}}^{\mathrm{ib}}$-2H and  $V_{\mathrm{Ga}}^{\mathrm{ic}}$-2H) the charge state is predicted as $-1$ \cite{varley_splitvacancy,varley_proton}.

\begin{figure*}[htbp]

    \begin{subfigure}[b]{0.22\linewidth}
        \includegraphics[width=\linewidth]{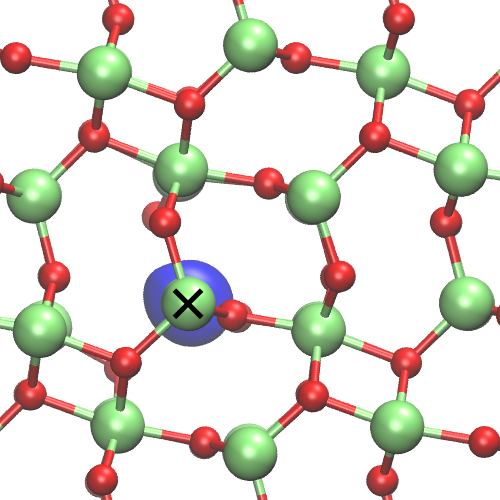}
        \caption{$V_{\mathrm{Ga1}}$}
        \label{f:vga1}
    \end{subfigure}
\hspace{0.3cm}
    \begin{subfigure}[b]{0.22\linewidth}
        \includegraphics[width=\linewidth]{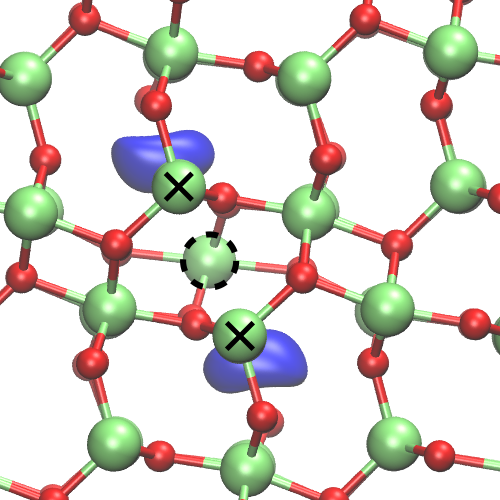}
        \caption{$V_{\mathrm{Ga}}^{\mathrm{ib}}$}
        \label{f:vib}
    \end{subfigure}
\hspace{0.3cm}
    \begin{subfigure}[b]{0.22\linewidth}
        \includegraphics[width=\linewidth]{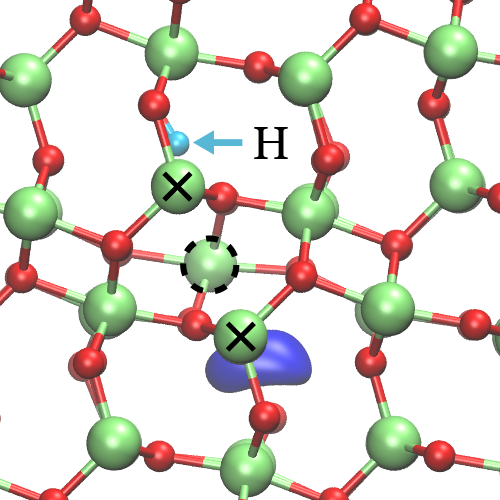}
        \caption{$V_{\mathrm{Ga}}^{\mathrm{ib}}$-1H}
       \label{f:vib-1h}
    \end{subfigure}
\hspace{0.3cm}
    \begin{subfigure}[b]{0.22\linewidth}
        \includegraphics[width=\linewidth]{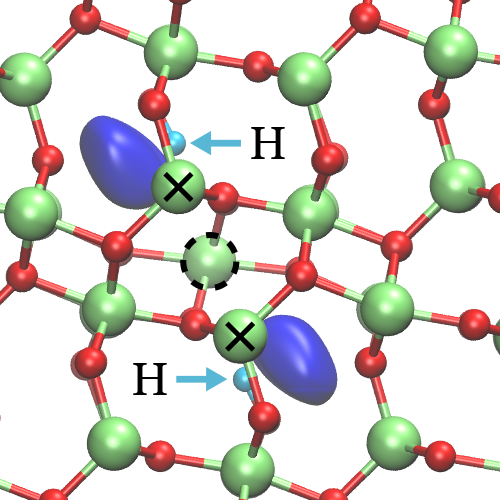}
        \caption{$V_{\mathrm{Ga}}^{\mathrm{ib}}$-2H}
        \label{f:vib-2h}
    \end{subfigure}
\hfill \hfill

\begin{subfigure}[b]{0.22\linewidth}
        \includegraphics[width=\linewidth]{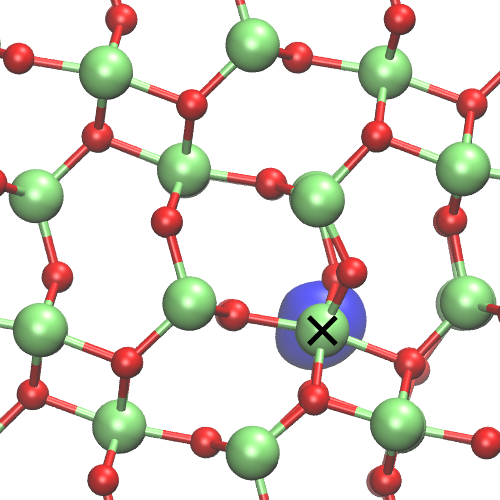}
        \caption{$V_{\mathrm{Ga2}}$}
        \label{f:vga2}
    \end{subfigure}
\hspace{0.3cm}
    \begin{subfigure}[b]{0.22\linewidth}
        \includegraphics[width=\linewidth]{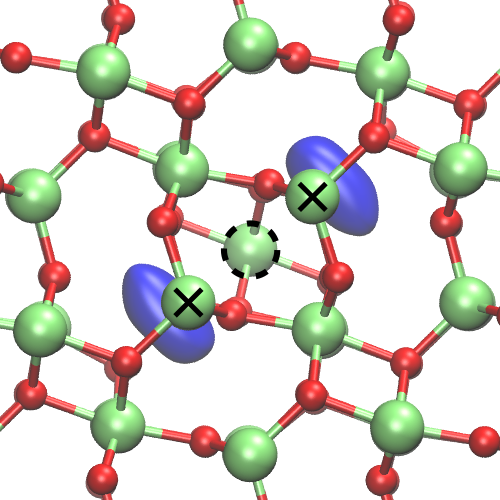}
        \caption{$V_{\mathrm{Ga}}^{\mathrm{ic}}$}
        \label{f:vic}
    \end{subfigure}
\hspace{0.3cm}
        \begin{subfigure}[b]{0.22\linewidth}
        \includegraphics[width=\linewidth]{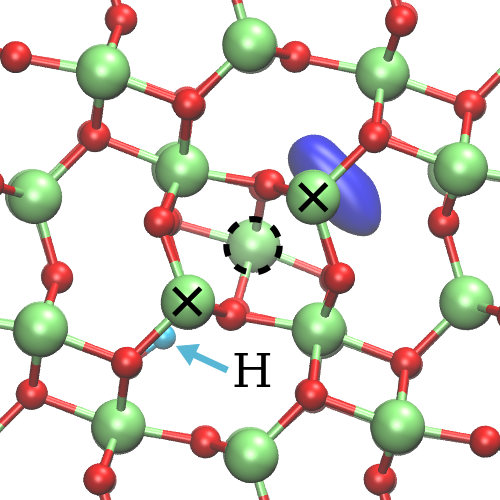}
        \caption{$V_{\mathrm{Ga}}^{\mathrm{ic}}$-1H}
       \label{f:vic-1h}
    \end{subfigure}
\hspace{0.3cm}
    \begin{subfigure}[b]{0.22\linewidth}
        \includegraphics[width=\linewidth]{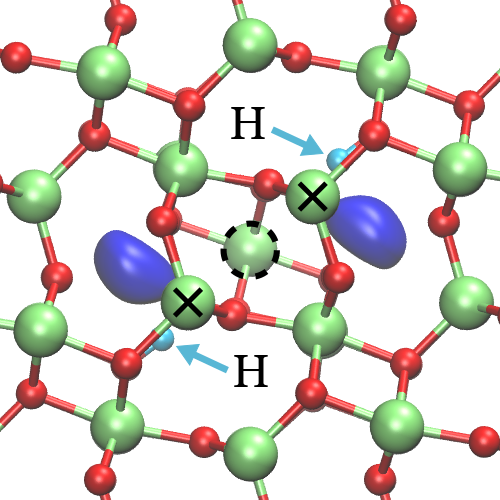}
        \caption{$V_{\mathrm{Ga}}^{\mathrm{ic}}$-2H}
       \label{f:vic-2h}
    \end{subfigure}
\hfill \hfill

\begin{subfigure}[b]{0.22\linewidth}
        \includegraphics[width=\linewidth]{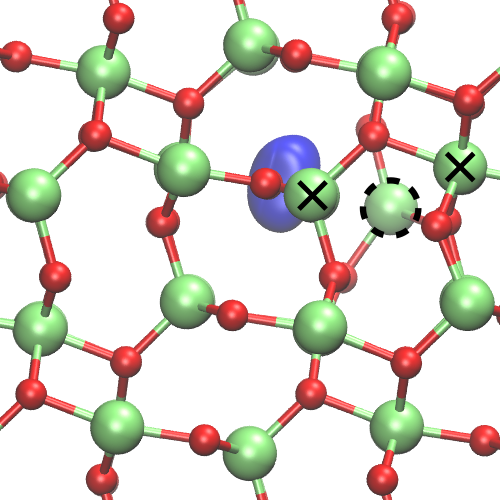}
        \caption{$V_{\mathrm{Ga}}^{\mathrm{ia}}$}
        \label{f:via}
    \end{subfigure}
\hspace{0.3cm}
\begin{minipage}[b]{0.22\linewidth}
\begin{center}
        \begin{subfigure}[t]{0.6\linewidth}
        \includegraphics[width=\linewidth]{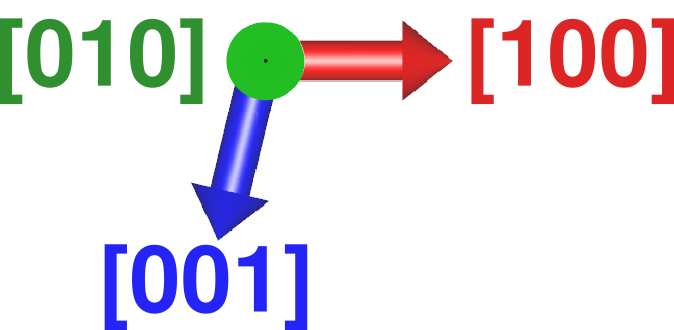}

    \end{subfigure}
    \vspace{0.2cm}     \vspace{0.2cm}
        \begin{subfigure}[t]{0.55\linewidth}
        \includegraphics[width=\linewidth]{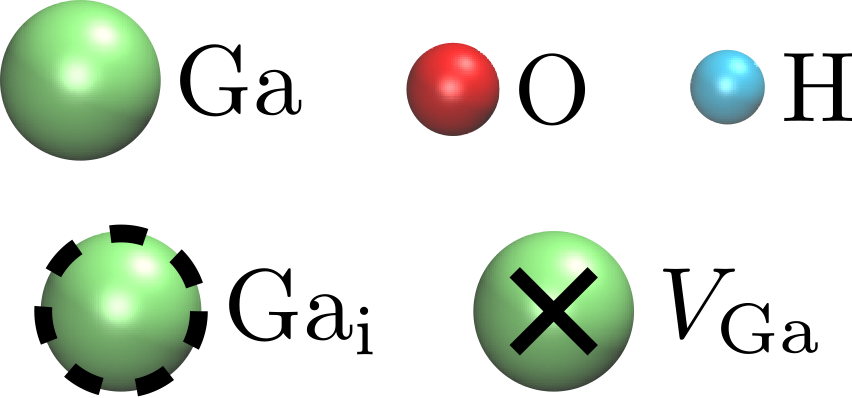}

    \end{subfigure}
\end{center}
    \vspace{1cm}     
    \end{minipage}
\hspace{0.3cm}
\begin{minipage}[b]{0.45\linewidth}
    \caption{Positron density isosurfaces (dark blue) in vacancies shown in the plane perpendicular to the [010] lattice vector (spanned by the [100] and [001] lattice vectors) \cite{vmd, tachyon}. The isosurfaces represent 63~\% of the maximum positron density.}
    \label{f:positron_densities}
\vspace{0.7cm} \vspace{0.7cm}
\vfill \vfill
\end{minipage}
\hfill     \hfill     

\end{figure*}

The calculated shapes of the density of the localized positron state can be also divided into three groups. The regular vacancies ($V_{\mathrm{Ga1}}$ and $V_{\mathrm{Ga2}}$) exhibit a sphere-like positron density as is typical of large enough vacancy defects, as illustrated in Fig. \ref{f:positron_densities}. In the case of vacancies with two symmetric open volumes, $V_{\mathrm{Ga}}^{\mathrm{ib}}$, $V_{\mathrm{Ga}}^{\mathrm{ic}}$, $V_{\mathrm{Ga}}^{\mathrm{ib}}$-2H and $V_{\mathrm{Ga}}^{\mathrm{ic}}$-2H, the positron localizes in the calculation equally into both half-vacancies. The positron density bodies of two symmetric half-vacancies are connected with a bridge of positron density with \SI{5}{\percent} of the maximum density value. Hence the half-vacancies can be considered as being part of the same potential well (and not two separate indistinguishable potential wells). In split Ga vacancies where the symmetry is broken, $V_{\mathrm{Ga}}^{\mathrm{ia}}$ (different Ga sites), $V_{\mathrm{Ga}}^{\mathrm{ib}}$-1H or $V_{\mathrm{Ga}}^{\mathrm{ic}}$-1H, the positron localizes into the larger open volume. In the case of $V_{\mathrm{Ga}}^{\mathrm{ib}}$-1H and $V_{\mathrm{Ga}}^{\mathrm{ic}}$-1H, the hydrogen occupies one of the half-vacancies and the positron localizes into the other half-vacancy. The positron density has a similar shape in these single-hydrogen half-vacancies as in the respective split Ga vacancies without hydrogen. In the split Ga vacancies with 2 hydrogen atoms, $V_{\mathrm{Ga}}^{\mathrm{ib}}$-2H and $V_{\mathrm{Ga}}^{\mathrm{ic}}$-2H, the hydrogen atoms occupy both "half-vacancies" making the situation symmetric again. As the positive charge of the hydrogen core occupies the center of the open volume it pushes the positron away from the center, changing the location and shape of the positron density compared to the  split Ga vacancies without hydrogen. 

\subsection{Positron lifetimes}
 
The calculated measurable quantities, the positron lifetimes (Table \ref{t:dft}) and Doppler broadening signals, follow roughly the same division of shape of the positron densities described in the previous section. 
The "regular" vacancies ($V_{\mathrm{Ga1}}$ and $V_{\mathrm{Ga2}}$) have the largest open volume and exhibit longest positron lifetimes, \SI{54}{\pico\second} above the calculated positron lifetime of 135 ps in the \gao\ lattice. Due to historical naming conventions we denote the positron lifetime in the lattice as $\tau_{\mathrm{B}}$ where the subscript B refers to "bulk". It should be noted that the predictive power of the state-of-the-art theoretical calculations in terms of the absolute scale of positron lifetimes (and ($S,W$) parameters) is low due to the different choices of approximations resulting in a wide range of, e.g., lattice lifetimes. Instead, differences between localized states and the lattice state can be compared in great detail between experiment and theory \cite{tuomistomakkonen}. 

In split Ga vacancies $V_{\mathrm{Ga}}^{\mathrm{ia}}$, $V_{\mathrm{Ga}}^{\mathrm{ib}}$ and $V_{\mathrm{Ga}}^{\mathrm{ic}}$,  the remaining open volume is smaller and the positron lifetime is $25-36$~ps longer than that in the lattice. The addition of a single hydrogen atom to $V_{\mathrm{Ga}}^{\mathrm{ib}}$ or $V_{\mathrm{Ga}}^{\mathrm{ic}}$ makes the positron to localize in the empty half of the split-vacancy without major changes to the positron density distribution, and the positron lifetime for $V_{\mathrm{Ga}}^{\mathrm{ib}}$-1H and $V_{\mathrm{Ga}}^{\mathrm{ic}}$-1H is essentially the same as for the same defects without hydrogen. Adding a second hydrogen atom significantly reduces the open volume and the resulting positron lifetimes for $V_{\mathrm{Ga}}^{\mathrm{ib}}$-2H and $V_{\mathrm{Ga}}^{\mathrm{ic}}$-2H are only $11-15$~ps longer than $\tau_{\mathrm{B}}$.

\begin{table}[htb]
  \caption{Calculated positron lifetimes and Doppler broadening signal anisotropies. The anisotropy spans are calculated dividing the maximum and minimum ($S,W$) parameters by the respective parameter of the $\beta$-Ga$_2$O$_3$ lattice in the [001] direction.}
  \label{t:dft}
\begin{center}
\setlength{\tabcolsep}{8pt}
\begin{tabular}{lrcc}
\hline
\hline
& \multicolumn{1}{c}{Positron} & \multicolumn{2}{c}{Anisotropy in}\\ System & lifetime (ps)&  S & W \\ 
 \hline 
Lattice & 135 & 1.000 - 1.020 & 0.97 - 1.00\\
$V_{\mathrm{Ga1}}$ & $\tau_{\mathrm{B}}$ +	54 & 1.018 - 1.040 & 0.90 - 0.97\\
$V_{\mathrm{Ga2}}$ & $\tau_{\mathrm{B}}$ +	54 & 1.022 - 1.038 & 0.91 - 0.96 \\
$V_{\mathrm{Ga}}^{\mathrm{ia}}$ & $\tau_{\mathrm{B}}$ +	25 & 1.005 - 1.023 & 0.95 - 0.99\\
$V_{\mathrm{Ga}}^{\mathrm{ib}}$ & $\tau_{\mathrm{B}}$ +	32 & 0.998 - 1.032 & 0.93 - 1.02\\
$V_{\mathrm{Ga}}^{\mathrm{ib}}$-1H & $\tau_{\mathrm{B}}$ +	27 & 0.991 - 1.031 & 0.92 - 1.05 \\
$V_{\mathrm{Ga}}^{\mathrm{ib}}$-2H & $\tau_{\mathrm{B}}$ +	11 & 1.002 - 1.023 & 0.93 - 0.99\\
$V_{\mathrm{Ga}}^{\mathrm{ic}}$ & $\tau_{\mathrm{B}}$ +	36 & 0.998 - 1.031 & 0.92 - 1.03\\
$V_{\mathrm{Ga}}^{\mathrm{ic}}$-1H & $\tau_{\mathrm{B}}$ +	32 & 0.986 - 1.031 & 0.92 - 1.08\\
$V_{\mathrm{Ga}}^{\mathrm{ic}}$-2H & $\tau_{\mathrm{B}}$ +	15 & 0.999 - 1.026 & 0.92 - 1.00 \\
$V_{\mathrm{O1}}$ & \multicolumn{3}{c}{does not trap positrons} \\
$V_{\mathrm{O2}}$ & \multicolumn{3}{c}{does not trap positrons} \\
$V_{\mathrm{O3}}$ & \multicolumn{3}{c}{does not trap positrons} \\
\hline
\hline
\end{tabular}
\end{center}
\end{table}

\subsection{Doppler broadening signals}

\begin{figure}[tbp]
\begin{center}
\includegraphics{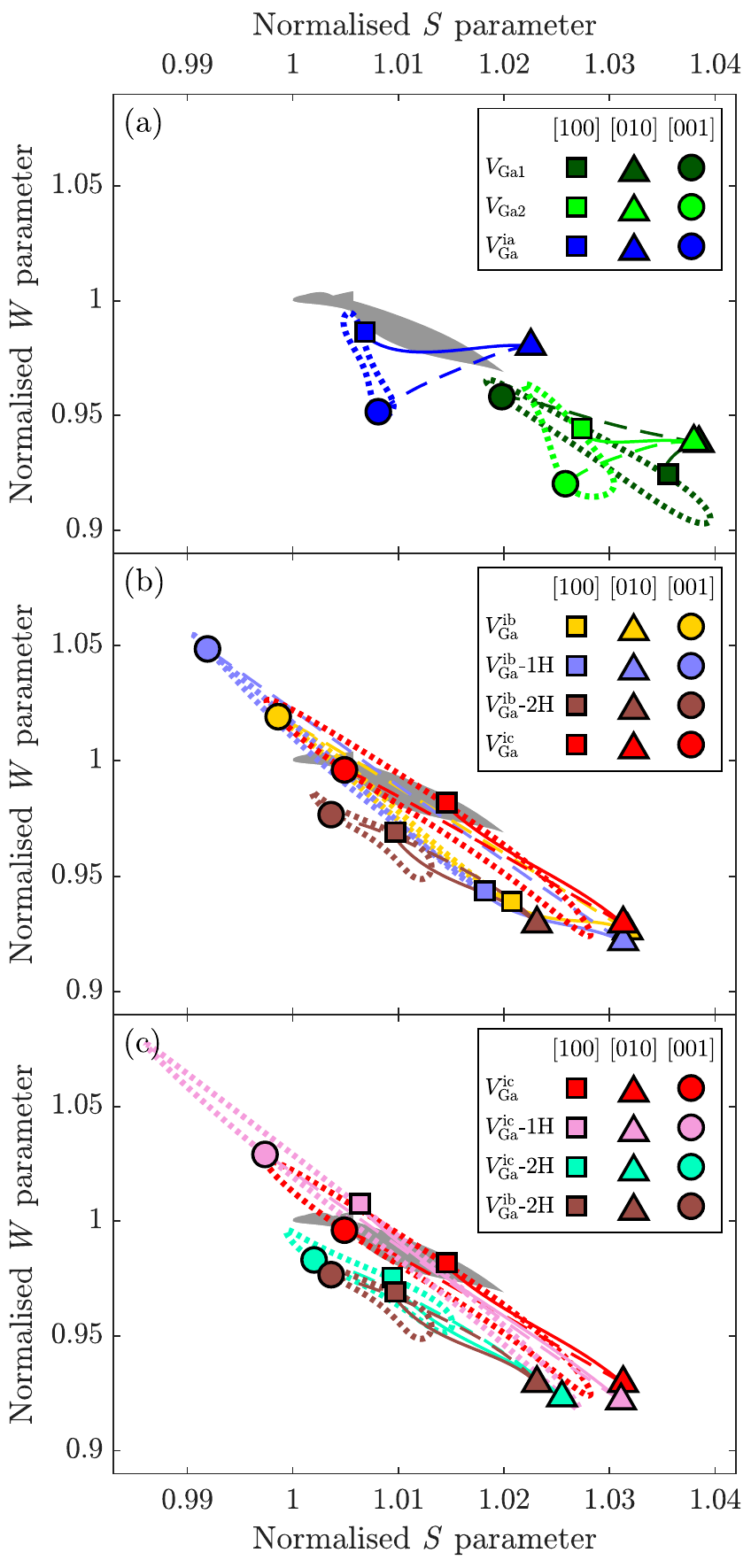}
\caption{The calculated ($S,W$) parameters in vacancies normalized to ($S,W$) parameters of $\beta$-Ga$_2$O$_3$ lattice in the [001] lattice direction. The grey shadow illustrates the ($S,W$) parameters of the $\beta$-Ga$_2$O$_3$ lattice from Fig. \ref{f:bulk_anisotropy}. The same notation is used for full, dashed and dotted curves to show ($S,W$) parameters on the geodesics between [100] and [010], [010] and [001], and [001] and [100], respectively.}

\label{f:dft_sw_tiled}
\end{center}
\end{figure}

Figure \ref{f:dft_sw_tiled} shows the calculated ($S,W$) parameters of the considered defects, normalized by the ($S,W$) parameters of $\beta$-Ga$_2$O$_3$ in the [001] lattice direction (the direction of the smallest calculated $S$ parameter in the $\beta$-Ga$_2$O$_3$ lattice). The figure uses the notation introduced in Fig. \ref{f:bulk_anisotropy} for square, triangle and circle to represent [100], [010] and [001] lattice directions, respectively, and full, dashed and dotted curves to represent ($S,W$) parameters on the geodesics between the [100] and [010], [010] and [001], and [001] and [100] directions, respectively. The  grey shadow illustrates the ($S,W$) of the $\beta$-Ga$_2$O$_3$ lattice (from Fig.~\ref{f:bulk_anisotropy}). The $\beta$-Ga$_2$O$_3$ lattice and all the considered defects have their maximum $S$ parameters in the [010] direction, while the smallest $S$ parameter lies in the [100]-[001] plane, in most cases in the vicinity of the [001] lattice direction. The defect ratio curves for the different directions are shown in the Appendix. They exhibit the same general behavior as the ($S,W$) parameters and are not discussed here in more detail. Figure \ref{f:dft_sw_tiled} visualizes the anisotropy issue already evident from the ($S,W$) parameter values shown in Table \ref{t:dft}. All of the considered vacancy defects exhibit at least as large an anisotropy in the Doppler signals as the \gao\ lattice, in contrast to what has been experimentally observed and/or theoretically calculated for materials such as Si, GaN, and ZnO where vacancies exhibit similar or smaller magnitude of anisotropy as the lattice \cite{Si_vacancy_anisotropy,ZnO_anisotropy}. In addition, the ($S,W$) parameters of many of the defects completely overlap with those of the \gao~ lattice.

The most distinct shape of anisotropies is exhibited by $V_{\mathrm{Ga1}}$, $V_{\mathrm{Ga2}}$ and $V_{\mathrm{Ga}}^{\mathrm{ia}}$ as their collection of ($S,W$) parameters reminds a right-pointing triangle (Fig. \ref{f:dft_sw_tiled}). The $V_{\mathrm{Ga1}}$ and $V_{\mathrm{Ga2}}$ have the highest $S$ parameters of the monovacancy-sized defects and their $S$ parameters span roughly from 1.02 to 1.04, see Table \ref{t:dft}. The two regular mono-vacancies share the same ($S,W$) parameters in the [010] lattice direction, while in the [100] and [001] directions the anisotropy of their Doppler parameters has a different shape and the minimum and maximum $S$ parameters in the [100]-[001] plane are in different lattice directions. The shape of the anisotropy of $V_{\mathrm{Ga}}^{\mathrm{ia}}$ reminds $V_{\mathrm{Ga2}}$ but $V_{\mathrm{Ga}}^{\mathrm{ia}}$ has smaller $S$ and higher $W$ parameters and it almost fully overlaps with the $\beta$-Ga$_2$O$_3$ lattice, with the $S$ parameter spanning from 1.005 to 1.023.

A second recognizable group is formed by defects with anisotropy along the diagonal, namely the split Ga vacancies $V_{\mathrm{Ga}}^{\mathrm{ib}}$ and $V_{\mathrm{Ga}}^{\mathrm{ic}}$ together with their singly hydrogenated versions  $V_{\mathrm{Ga}}^{\mathrm{ib}}$-1H and $V_{\mathrm{Ga}}^{\mathrm{ic}}$-1H (see Figs. \ref{f:dft_sw_tiled}b and \ref{f:dft_sw_tiled}c). The calculated Doppler broadening results predict that in the [100]-[001] plane their ($S,W$) parameters swing all the way from the smallest $S$ parameter up to almost the maximum $S$ parameter in the [010] lattice direction. The $V_{\mathrm{Ga}}^{\mathrm{ib}}$, $V_{\mathrm{Ga}}^{\mathrm{ic}}$, $V_{\mathrm{Ga}}^{\mathrm{ib}}$-1H  and $V_{\mathrm{Ga}}^{\mathrm{ic}}$-1H have the same ($S,W$) parameters in the [010] lattice direction. The ($S,W$) parameters of the split Ga vacancies $V_{\mathrm{Ga}}^{\mathrm{ib}}$ and $V_{\mathrm{Ga}}^{\mathrm{ic}}$ (Fig \ref{f:dft_sw_tiled}b) span from 0.998 to 1.032 in $S$ and from 0.92 to 1.03 in $W$, covering the ($S,W$) range of the $\beta$-Ga$_2$O$_3$ lattice completely. They also overlap with each other very strongly. However, in the [100]-[001] plane $V_{\mathrm{Ga}}^{\mathrm{ib}}$ has the smallest $S$ parameter close to the [001] direction and maximum in [100] whereas $V_{\mathrm{Ga}}^{\mathrm{ic}}$ reaches similar ($S,W$) parameter values but in directions which are rotated \SI{45}{\degree} from the [100] lattice vector towards the [001] lattice vector. The ($S,W$) parameters in  $V_{\mathrm{Ga}}^{\mathrm{ic}}$ in the [100] and [001] lattice directions  are roughly in the middle of the total ($S,W$) parameter range of the [100]-[001] plane. 

Adding one hydrogen to the split Ga vacancies $V_{\mathrm{Ga}}^{\mathrm{ib}}$ and $V_{\mathrm{Ga}}^{\mathrm{ic}}$ localizes the positron to only one of the half-vacancies without dramatically changing the shape of the positron density and with only a minor effect on the positron lifetime. In Doppler broadening results, the addition of one hydrogen increases the anisotropy span but keeps the ($S,W$) parameters of the [010] lattice direction intact. The ($S,W$) parameters of the [100] and [001] lattice directions, and the whole [100]-[001] plane, are spread towards smaller $S$ and larger $W$ parameters. $V_{\mathrm{Ga}}^{\mathrm{ic}}$-1H is found to have the smallest $S$ parameter (0.986), significantly smaller than that in the $\beta$-Ga$_2$O$_3$ lattice, and the $S$ parameters of $V_{\mathrm{Ga}}^{\mathrm{ib}}$-1H and $V_{\mathrm{Ga}}^{\mathrm{ic}}$-1H span from 0.991 to 1.031. Adding a second hydrogen atom changes the shape of anisotropy of $V_{\mathrm{Ga}}^{\mathrm{ib}}$-2H and $V_{\mathrm{Ga}}^{\mathrm{ic}}$-2H and they become almost indistinguishable. Their shape of anisotropy is somewhat similar to the $\beta$-Ga$_2$O$_3$ lattice but they exhibit slightly lower $W$ parameter values. The whole range of the anisotropy is still much larger than in the lattice, and the $S$ parameter spans roughly from 1.00 to 1.025.

\section{Anisotropy of the electron-positron momentum density}

The momentum density of annihilating pairs measured in a Doppler broadening experiment reflects the anisotropy present in the ionic and electronic structures of the lattice. It can be viewed as the electron momentum density "as seen by the positron". The possible anisotropy of the annihilation signal of the delocalized lattice state is affected by a number of factors, including:
\begin{enumerate}
    \item The electron momentum density of the host lattice (in the absence of any positrons, as in X-ray Compton scattering experiments \cite{MakkonenJPCS2005}). The anisotropy of the lattice itself can result in an anisotropic electron momentum density.
    \item The momentum density of the positron. Confinement in one or more directions can lead to a corresponding broadening.
    \item The positron density distribution in the lattice and its overlap with the electronic orbitals. In the delocalized state in a defect-free lattice, the positron probes the interstitial region far away from the repulsive nuclei. When localized at a vacancy in \gao\ the annihilation occurs mainly with valence electron states of the neighboring atoms. Both Ga 3d and O 2s/p states have a rather broad momentum distribution.
    \item Electron-positron correlation effects.
\end{enumerate}
    
The \gao\ structure has a low symmetry compared to many well-known systems with anisotropic positron signals such as Si in the diamond lattice structure. As noted above, the positron density distribution is rather anisotropic and runs along the tubes in the [010] direction. In Si the positrons also favor specific channels along the  \{110\} directions (see Fig.~\ref{f:si_latticestate}) but the intersecting tubes and the annihilating pair momentum density retain the cubic symmetry of the lattice. On the other hand, point defects in \gao\ occur only in specific orientations (see, for example, Fig.~\ref{f:vacancy_structure}) and their anisotropic fingerprints do not average out to a more isotropic spectrum. This can be compared to, for example, a vacancy-donor pair in Si that has 4 possible symmetric orientations occurring randomly in a real sample. In conclusion, for single crystals, in which the positron signal is not averaged between differently oriented grains, the anisotropy of the signal of any \gao\ sample naturally follows from the structure of the lattice and the simple point defects.

For a positron delocalized in the \gao\ lattice, we have analyzed the points 1 and 2 of the above list in detail. First, we have calculated the electron momentum density and the Compton profiles within the independent-particle model and impulse approximation in full consistence with our positron modeling. This is equivalent to setting the positron orbital to a constant and neglecting the enhancement factors in the model \cite{AlataloPRB1996} we use for the momentum density of annihilating pairs. Second, we analyze the positron momentum density (the Fourier coefficients of the orbital) in order to understand if the anisotropy of the positron orbital alone could increase the anisotropy of the momentum density of annihilating pairs. The Compton profile lineshapes ($S$ and $W$ type parameters extracted from the profiles) turn out to reflect the same kind of anisotropy as the Doppler spectra in the \gao\ lattice, which implies that the lattice state anisotropy is an inherent property of the \gao\ electronic structure. According to the Fourier components of the positron orbital, the positron is "free" along the [010] direction and confined in the perpendicular directions (see Fig.~\ref{f:si_latticestate}), as in the perpendicular directions the positron momentum density displays a broadening. The anisotropic nature of the positron momentum density has a role in determining the full anisotropy, as the momentum density of annihilating pairs is approximately the electron momentum density convoluted with the positron momentum density. Also, the positron density gives a larger weight to the outermost valence orbitals of ions in the interstitial region of the lattice, where the electron-positron correlations have the strongest effects. Separating and quantifying the roles of these mechanisms is, however, difficult.

In the case of defects trapping the positron, the Doppler broadening measures the electron-momentum density at the annihilation site but the confinement of the positron might also play a role. We find that the shape of the positron density is correlated to the ($S,W$) parameters in \gao\ in an intriguing way. In the \gao\ lattice and defects, the overall highest $S$ parameters (and typically the lowest $W$ parameters) are found in the [010] lattice direction. This direction corresponds to the open structure and the least dense lattice planes (Fig.~\ref{f:structure}) and to the direction along which the tubular positron states are formed in the \gao\ lattice (Fig.~\ref{f:latticestates}). Interestingly, in the [100]-[001] plane the local extrema of the ($S,W$) parameters are not found in "high-symmetry" [100] and [001] lattice directions in all cases. This is particularly visible in  $V_{\mathrm{Ga}}^{\mathrm{ib}}$ and $V_{\mathrm{Ga}}^{\mathrm{ic}}$ that are very similar defects: their structures differ mainly by a rotation of $\sim$\SI{105}{\degree} (Fig. \ref{f:vacancy_structure}) and their positron lifetimes are essentially the same (Table \ref{t:dft}). However, the directions of the local maxima and minima of the ($S,W$) parameters in the [100]-[001] lattice plane differ by $\sim$\SI{45}{\degree}.

The directions of the local ($S,W$) parameter extrema in the [100]-[001] plane are correlated with the direction of the longitudinal axis of the the positron density in all defects where the positron density has a clearly non-spherical shape. The minima of the $S$ parameter in the [100]-[001] plane are rotated by \SI{90}{\degree} from the in-plane $S$ parameter maxima, and hence correlated with the "narrow axis" of the positron density. These correlations are best visible in $V_{\mathrm{Ga}}^{\mathrm{ib}}$ and $V_{\mathrm{Ga}}^{\mathrm{ic}}$, and in particular when adding a second hydrogen to $V_{\mathrm{Ga}}^{\mathrm{ib}}$. The maximum $S$ parameter in the [100]-[001] plane in $V_{\mathrm{Ga}}^{\mathrm{ib}}$ and $V_{\mathrm{Ga}}^{\mathrm{ib}}$-1H is in the [100] direction while the longitudinal axis of their positron densities point is aligned with the [100] lattice direction as well. The maximum $S$ parameter of $V_{\mathrm{Ga}}^{\mathrm{ic}}$, $V_{\mathrm{Ga}}^{\mathrm{ic}}$-1H and $V_{\mathrm{Ga}}^{\mathrm{ic}}$-2H in the [100]-[001] plane is found rotated by \SI{45}{\degree} from [100] towards the [001] lattice vector, as does the longitudinal axes of their positron densities. Adding a second hydrogen to $V_{\mathrm{Ga}}^{\mathrm{ib}}$-2H rotates the positron density by approximately \SI{45}{\degree} from the [100] lattice vector towards [001], parallel to positron densities in the $V_{\mathrm{Ga}}^{\mathrm{ic}}$ defects. The local ($S,W$) parameter extrema for $V_{\mathrm{Ga}}^{\mathrm{ib}}$-2H are in the same directions as for $V_{\mathrm{Ga}}^{\mathrm{ic}}$-2H.

The correlation between the shape of the positron density and the Doppler broadening parameters is consistent with the following mechanism. An elongated positron density in a certain direction indicates less localization in this spatial dimension. Less localization in real space goes hand-in-hand with a stronger localization in momentum space, that is a narrower momentum distribution. However, we point out that also the local ionic structure and its orientation plays a role. In any case, for localized positrons the overall anisotropy of the Doppler signals increases by a factor of up to 2-3 in the \textit{ib} and \textit{ic} type split Ga vacancies. 

Another manifestation of large anisotropy in positron annihilation radiation of a reduced symmetry system is the case of graphite~\cite{TangPRB2002}, in which the positron is confined in 2D states between the sheets and samples predominantly the $p_{z}$ orbitals of the carbon atoms, giving rise to a similar bimodal structure in the momentum density of annihilating pairs and strongly anisotropic Doppler spectra. This comparison to graphite demonstrates that the unusual magnitude of the anisotropic features in the Doppler broadening in \gao\ is only colossal when compared to typical widely studied semiconductors such as Si, GaAs, GaN or ZnO with high-symmetry crystal structures. Two-dimensional \cite{TangPRB2002} and in the case of \gao\ one-dimensional positron states should perhaps be expected to produce anisotropic Doppler broadening signals, and the exact nature of the positron state in a given crystal structure can only be determined by performing advanced theoretical calculations. Finally, it should be noted that the strongly one-dimensional positron state in the $\beta$-Ga$_2$O$_3$ lattice suggests that positron diffusion might be significantly faster along the [010] lattice direction than in the other directions. This should be considered in detail in future experiments.

\section{Experimental anisotropy}

\subsection{Sample orientation in experiments}

We compare the results of our theoretical calculations to experimental results obtained in two semi-insulating single crystal \gao\ bulk samples (S1 and S2 in the following) that are grown by the Czochralski method and doped with Mg (described in more detail in Ref. \onlinecite{galazka}). The surface of the plate-like sample S1 is in the (100) crystal plane and the surface of S2 in the (010) plane. These single crystal samples are identical to those used as substrates in \gao\ thin film growth and exhibit the same measurement orientation dependence as the $n$-type thin films as shown in Ref.~\onlinecite{spie}. To exclude possible experimental artefacts, the Doppler broadening of the positron-electron annihilation radiation was measured in the samples in two different ways, with a slow positron beam and with a fast positron setup.

The fast positron measurements were performed on two identical pieces of S1 in the three lattice directions of the standard conventional unit cell of $\beta$-Ga$_2$O$_3$, and along the geodesics connecting the lattice directions with a \SI{10}{\degree} step. In the fast positron setup, a high-purity Ge (HPGe) detector with energy resolution of \SI{1.15}{\kilo\electronvolt} at \SI{511}{\kilo\electronvolt} was used to record the annihilation photons emitted from two sample pieces with a positron source sandwiched in between. The positron source with 1~MBq of activity was composed of $^{22}$Na encapsulated in \SI{1.5}{\micro\meter} thick Al-foil. The amount of source annihilations was determined  with positron lifetime measurements to be less than \SI{4}{\percent}. These lifetime measurements also reveal that the crystals only show a single lifetime component of $\sim 185$~ps at room temperature. Based on earlier reports on \gao\ single crystals the bulk lifetime should be at most $175-180$~ps \cite{korhonen2015,ting_mseb_2002}, suggesting that these crystals contain vacancy-type defects with relatively short lifetime components compared to the bulk lifetime as they are unresolvable. We do not discuss the lifetime results in more detail in this work, but report that the experiments were performed with a standard digital spectrometer
in collinear geometry and a Gaussian time resolution of 250 ps (FWHM). The distance between the sample-source sandwich and the detector was \SI{35}{\centi\meter} yielding an angular resolution of about  \SI{10}{\degree}  as defined by the solid angle covered by the detector crystal. The crystal orientations of the samples were determined by X-ray diffraction measurements. The sample-source sandwich was rotated to collect annihilation spectra in all desired directions. The ($S,W$) parameter windows were set as $0 - 0.45$~a.u. ($0 - 0.83$~keV) for the $S$ and to $1.54 - 4.07$~a.u. ($2.87- 7.60$~keV) for the $W$ parameter. A total of $10^6$ counts was recorded for each spectrum. The background was subtracted by taking into account the effect of higher-energy annihilation events, as discussed in Ref. \onlinecite{krause1999positron}. 

In the slow positron beam experiments, two unspecified perpendicular directions within the surface plane of the sample have been measured in samples S1 and S2 with two HPGe detectors that have an energy resolution of \SI{1.25}{\kilo\electronvolt} at \SI{511}{\kilo\electronvolt}, as reported in Ref. \onlinecite{spie}. In S1, the measurement directions were in the plane spanned by the [010] and [001] lattice vectors, while for S2, the measurements were performed in the plane spanned by the [100] and [001] lattice vectors. The aspect ratio is clearly different from the fast positron setup as the detectors are only at a distance of roughly 3~cm from the sample. The ($S,W$) parameter values were acquired at positron implantation energy of \SI{25}{\kilo\electronvolt}, corresponding to mean stopping depth \SI{1.2}{\micro\meter}, so that the back-diffusion generated annihilations at the surface do not affect the data. The same ($S,W$) parameter windows were used in the data analysis as in the fast positron experiments.

\begin{figure}[htb]
  \begin{center}

    \begin{subfigure}[t]{\linewidth}
      \begin{center}
        \includegraphics{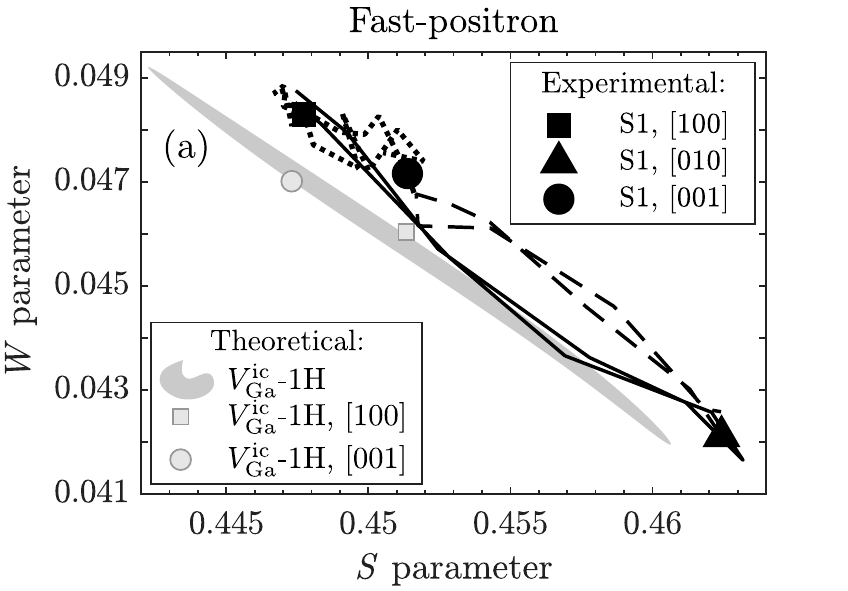}

      \end{center}
    \end{subfigure}

    \vspace{0.2cm}
    
    \begin{subfigure}[t]{\linewidth}
      \begin{center}

        \includegraphics{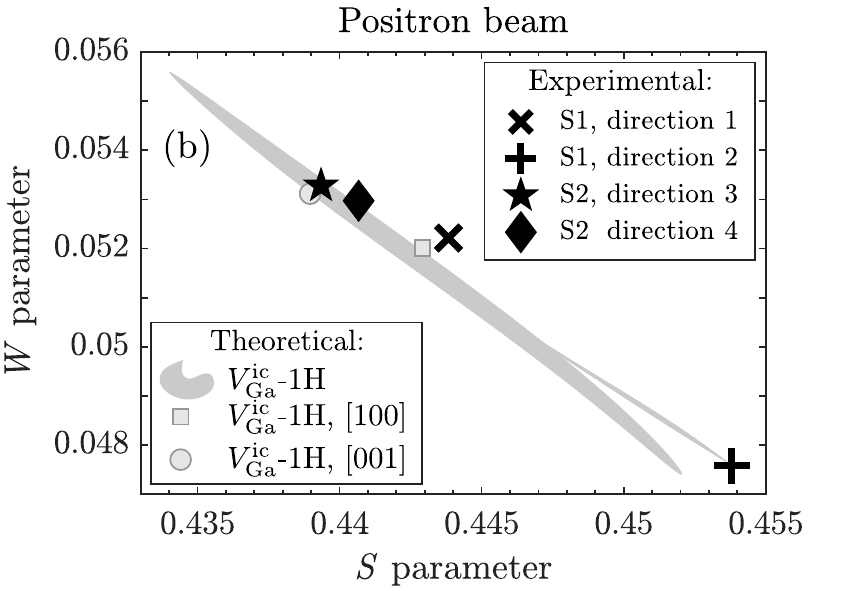}

      \end{center}
    \end{subfigure}
    \caption{Experimental ($S,W$) parameters of the studied \gao\ samples. The grey shadow illustrates the calculated ($S,W$) parameters of $V_{\mathrm{Ga}}^{\mathrm{ic}}$-1H, the defect with the largest anisotropy, as shown in Fig. \ref{f:dft_sw_tiled}.  
(a) Standard fast positron measurement of sample S1 in the lattice vector directions and on the geodesics between them with the notation introduced in Fig. \ref{f:bulk_anisotropy}. 
(b) Sample S1 (100) and S2 (010) measured with a slow positron beam at an acceleration voltage of 25 kV. Two  directions perpendicular to the sample surface normal were measured for both samples.}
    \label{f:exp}
  \end{center}
\end{figure}

The results of the fast positron measurements are shown in Fig. \ref{f:exp}a. The figure uses the notation introduced in Fig. \ref{f:bulk_anisotropy}, and accordingly, the lines in Fig. \ref{f:exp}a describe the rotations with \SI{10}{\degree} step. The experimental ($S,W$) parameter values in the lattice directions [100], [010] and [001], and along the geodesics between them, behave almost linearly along the diagonal in the figure. This is similar to the calculated behavior of the $(S,W)$ parameters in the \gao\ lattice and the \textit{ib} and \textit{ic} type split Ga vacancies, and clearly distinct from the standard Ga vacancies and the \textit{ia} type split Ga vacancy. The ($S,W$) parameters of the [100] and [001] directions are rather close to each other while the [010] direction has clearly the highest $S$ parameter. The positron beam measurement shows features similar to the fast positron measurement (Fig. \ref{f:exp}b). It is natural that the absolute ($S,W$) parameter values differ from the fast positron measurement: these two experiments were performed in different setups with different detectors and measurement geometries. In addition, in the fast positron measurement there is a $\lesssim$\SI{4}{\percent} isotropic source contribution to the annihilations. The measurement directions 1 and 2 of sample S1 share a close resemblance to the [100] and [010] lattice directions of the fast positron experiment, respectively. The ratio of the ($S,W$) parameters of directions 1 and 2 are 1.023 for the $S$ and 0.90 for the $W$ parameter, while the corresponding ratios of [100] and [010] crystal directions in the fast positron measurement are 1.025 and 0.90, respectively. This suggests that the directions 1 and 2 of sample S1 are close to [100] and [010] lattice directions. In sample S2, directions 3 and 4 differ less than [100] and [001] suggesting that the measurement directions are slightly off from [100] and [001] lattice directions.

The overall magnitude of the anisotropy observed in the positron beam measurement (S1 direction 2, S2 direction 3) is 1.033 for the $S$ and 0.89 for the $W$ parameter. This is similar in the fast positron measurement of sample S1, where the overall magnitude is 1.035 for the $S$ and 0.84 for the $W$ parameter. Here the anisotropy is defined as the largest (smallest) $S$ ($W$) parameter divided by the smallest (largest) $S$ ($W$) parameter. Figure \ref{f:exp} also shows the theoretically calculated ($S,W$) parameter of $V_{\mathrm{Ga}}^{\mathrm{ic}}$-1H, the defect with the largest overall anisotropy, with a grey shadow and grey markers along the lattice vectors. The lack of a "defect-free" reference \gao\ sample prevents from normalizing the experimental ($S,W$) parameters, and the comparison of normalized theoretical and experimental ($S,W$) parameters. Instead, we shifted the ($S,W$) parameters of the theoretical calculations so that the [010] lattice direction matches the maximum $S$ (and minimum $W$) parameter of the experiments. The operation does not affect the shape or relative magnitude of the anisotropy and allows us to compare the experimental and theoretical anisotropies. Clearly the scale of the anisotropy in the Doppler signals is the same in experiments and theory, and an order of magnitude higher than observed in other semiconductors \cite{Si_anisotropy, ZnO_anisotropy}, comparable to the anisotropy in graphite \cite{graphite}. Interestingly, the experimental anisotropy is comparable in magnitude to that in $V_{\mathrm{Ga}}^{\mathrm{ic}}$-1H, which is three times larger than that of the \gao\ lattice. Note that the defect predicted to have the lowest formation energy \cite{varley_splitvacancy}, namely the $V_{\mathrm{Ga}}^{\mathrm{ib}}$-2H, has an overall anisotropy that is twice that of the \gao\ lattice, but somewhat smaller than the experimental anisotropy. These observations suggest that the single crystal \gao\ samples contain high concentrations of (hydrogenated) split Ga vacancies. 

\begin{figure}[h]
  \centering
    \includegraphics{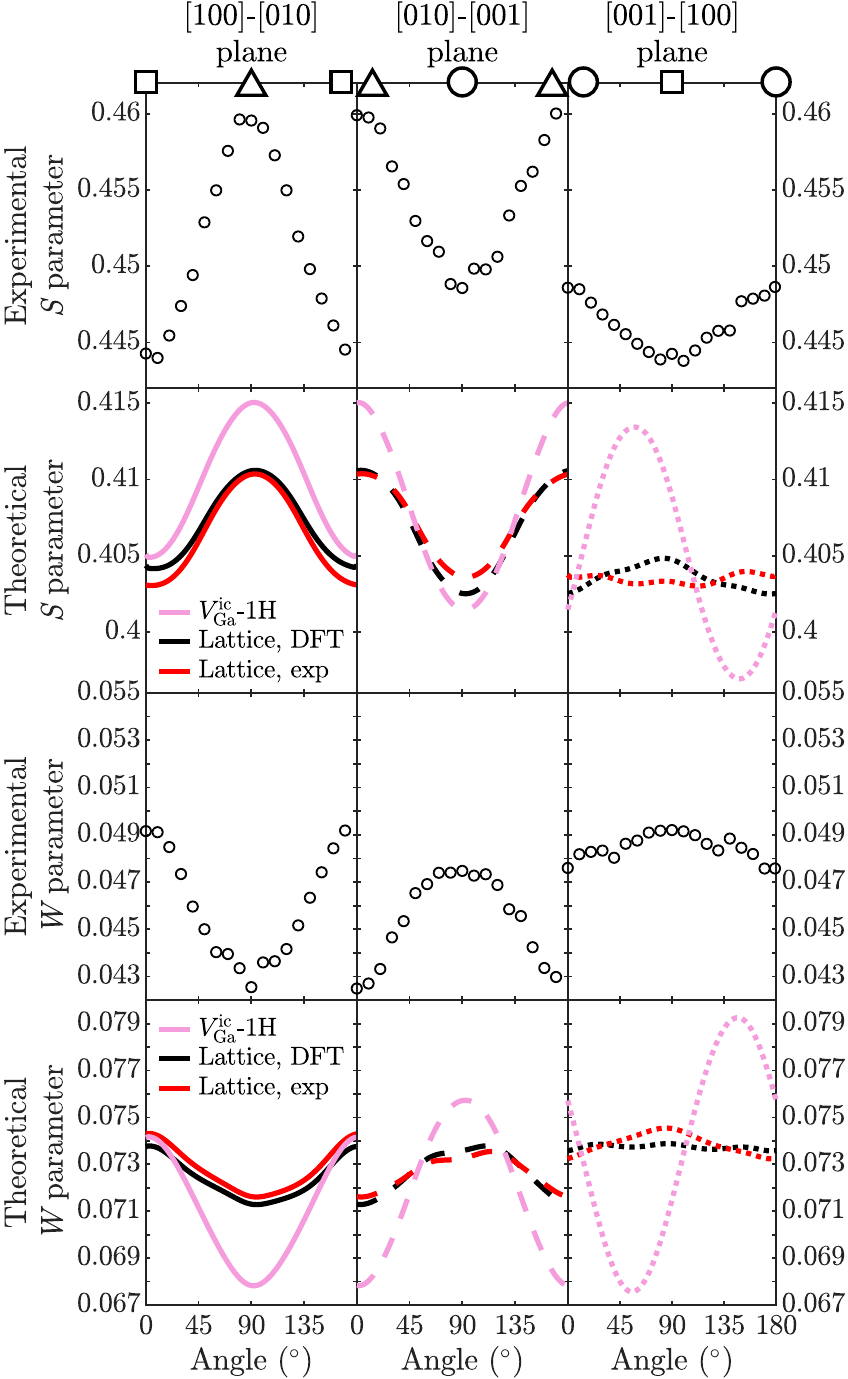}
        \caption{Experimental and theoretical ($S,W$) parameters as a function of measurement angle along different geodesics. Squares, triangles and circles at the top denote the [100], [010] and [001] lattice directions, respectively. The experimental data are obtained in sample S1 in the angle-resolved fast-positron experiment and the theoretical data consist of \vic-1H and the \gao\ lattice using theoretical and experimental  lattice structures (as described in Fig. \ref{f:bulk_sw}).}
         \label{f:angledependent}
     \end{figure}

Further comparison between theory and experiment is presented in Fig.~\ref{f:angledependent}, where we show $S$ and $W$ parameters as a function of the measurement orientation on the three geodesics. In addition to the calculated $V_{\mathrm{Ga}}^{\mathrm{ic}}$-1H
data, we also show the $S$ and $W$ parameters of the \gao\ lattice calculated using both the DFT and experimental unit cells discussed in Section~\ref{lattice}. Note that in this figure we show the calculated data without adjustment to match the [010] direction, but have set the ranges of vertical axes to be of equal magnitude for both the experimental and theoretical data to allow for visual comparison. The general behavior of both experimental and theoretical data along the geodesics in the [100]-[010] and [001]-[010] planes is the same. However, it is also evident that the magnitude of the anisotropy is much larger in the experiments than in the \gao\ lattice calculations, with a clearly better match to $V_{\mathrm{Ga}}^{\mathrm{ic}}$-1H. In the [100]-[001] plane the comparison is less straightforward. The overall magnitude in experiments and theory is a better match for both types of \gao\ lattice data, while the qualitative behavior in the experiments is a good match to the \gao\ lattice data calculated with the experimental unit cell. The calculated data for $V_{\mathrm{Ga}}^{\mathrm{ic}}$-1H are distinctly different from the experimental data. These observations suggest that the experimental signals, while containing a significant contribution from (hydrogenated) split Ga vacancies, also carry \gao\ lattice related contributions.

Our results show that from the point of view of the anisotropy of the Doppler signals, experiment and theory are in excellent agreement in three aspects: (i) the maximum $S$ (minimum $W$) are found in the [010] direction, (ii) the general shape of the anisotropy follows similar trends and (iii) the overall magnitude of the anisotropy is colossal compared to other studied 3D semiconductor crystals. Uncertainties in the experiment-theory comparison remain in the data obtained in the [100]-[001] plane. In both experiment and theory, the [100] and [001] directions are similar in the sense that they produce ($S,W$) parameters that are quite close to each other and clearly far away from the [010] direction. However, the calculations predict the $S$ parameter in the [001] lattice direction to be smaller than in the [100] direction for nearly all of the considered defects, while the positron experiments (including also our unpublished results) consistently show the [100] lattice direction to have lower $S$ parameter than the [001] direction. Further research is required to resolve this issue that could be related to the uncertainty in the exact atomic positions in the \gao\ lattice, as demonstrated by the different results obtained by employing differently determined unit cells in the modeling. Also, when directly comparing experimental and theoretical $(S,W)$ results resulting from rotating the sample, we have to stress that even though we take into account the experimental energy resolution of the detectors, our simulations do not yet account for the finite size of the detectors and integration over a finite solid angle especially in measurements made with slow positron beams. Including the effect of a proper angular resolution function in modeling can be expected to smoothen the behavior of the $(S,W)$ values in graphs such as Fig.~\ref{f:dft_sw_tiled} and limit the extrema especially if they occur only in specific directions covered by a small solid angle (see, for example, Fig.~\ref{f:bulk_s_sphere}).

\subsection{Defect identification}

\label{c:implications}

The strong overlap (see Fig.~\ref{f:dft_sw_tiled}) of the lattice and vacancy $(S,W)$ parameters in \gao\ requires that care is taken to measure all samples along the exact same lattice orientation in defect studies with Doppler broadening, a precaution that is not necessary in other semiconductors. While the identification of a "defect-free" reference \gao\ sample that would be known to produce the \gao\ lattice signals is still missing, the unusual anisotropy of the $(S,W)$ parameters of the lattice and the various kinds of vacancy defects opens another approach for defect identification in \gao. It is worth noting that in fast positron experiments the measurement orientation of the samples can be chosen freely, but in thin film studies with a slow positron beam the measurement setup is more restricted, typically allowing for signal acquisition either along the beamline (measurement orientation perpendicular to the sample surface) or in a fixed direction perpendicular to the beamline (measurement orientiation in the plane of the sample surface). In the latter case, measuring two perpendicular directions is a reasonable approach for utilizing the positron signal anisotropy in slow positron beam studies. Hence, the surface orientation of the samples is relevant for positron studies, and it is important to know the in-plane orientation as well in order to choose the appropriate measurement geometry. For the benefit of future studies, we have included in the Appendix figures that  \ref{f:sw_directions} illustrate how the normalized ($S,W$) of the \gao\ lattice and defects behave in the [001], [010] and [100] lattice directions.

Comparison of the experimental ($S,W$) parameters (Fig.~\ref{f:exp}) and calculated ($S,W$) parameters (Fig.~\ref{f:dft_sw_tiled}) shows that the magnitude of the experimental anisotropy is 1.035 for the $S$ and 0.84 for the $W$ parameter, almost double the $S$ anisotropy and 5-fold the $W$ anisotropy calculated for the $\beta$-Ga$_2$O$_3$ lattice. This strongly suggests that the experimental positron data obtained in the \gao\ single crystal is strongly affected by defect signals that show larger overall anisotropy, which means that the concentration of positron-trapping vacancy defects in the sample is at least $\sim 10^{18}$~cm$^{-3}$. We discard the possibility of a major contribution from $V_{\mathrm{Ga1}}$, $V_{\mathrm{Ga2}}$ and $V_{\mathrm{Ga}}^{\mathrm{ia}}$ in the experimental data as the shape of the anisotropy is clearly different. The shape of the experimental anisotropy is close to the shape of all the other calculated defects, that is split Ga vacancies of \textit{ib} and \textit{ic} type, with and without hydrogen. In the experimental results, the ($S,W$) parameters in the [100] and [001] lattice directions are much closer to each other than the parameters in the [010] direction, which suggests that the experimental data resemble more the $V_{\mathrm{Ga}}^{\mathrm{ic}}$ -type split Ga vacancies. However, as the calculated ($S,W$) parameters vary a lot along the geodesics connecting [100] and [001] in the $V_{\mathrm{Ga}}^{\mathrm{ib}}$-type split Ga vacancies, and due to the uncertainty in the proper unit cell (see Section \ref{lattice}), this is not as strong an argument as the one ruling out the first set of defects. The magnitude of the calculated anisotropy for the doubly hydrogenated split Ga vacancies is smaller than that in the experiments, suggesting that the dominant contribution to the experimental data could indeed come from the clean and singly hydrogenated split Ga vacancies. However, at this stage we cannot rule out the possible contribution of more complex defects such as split Ga vacancy – O vacancy complexes. The assignment of experimentally observed anisotropy of the Doppler signals to $V_{\mathrm{Ga}}^{\mathrm{ic}}$-related defects is also supported by the recent theoretical and experimental work: $V_{\mathrm{Ga}}^{\mathrm{ic}}$ has been predicted to have the lowest formation energy among clean mono-vacancy type defects \cite{varley_proton}, and the existence of $V_{\mathrm{Ga}}^{\mathrm{ic}}$ was observed with STEM \cite{stem}. The fact that these defects could be found by STEM agrees well with the estimate from positron annihilation: in both experiments, the split Ga vacancy concentration needs to be at least $10^{18}$~cm$^{-3}$.

It is worth considering the interpretations and discussion of the experimental results presented in Refs. \onlinecite{korhonen2015,spie} in the light of our present results. Most of the experimental data in those reports were measured with a slow positron beam in thin films grown on (100) oriented \gao\ substrates. After the publication of those reports, it was found that the measurements were performed in a "low-$S$" direction that probably is close to the [001] lattice direction (as the other direction in that plane is [010], a "high-$S$" direction). The ($S,W$) data in Refs. \onlinecite{korhonen2015,spie} fall on the dashed line (or its extension) presented in Fig.~\ref{f:sw_directions}a (see Appendix), if the "bulk" measurement data point is associated to any of the \text{ib} or \text{ic} type split Ga vacancies or the lattice. The data points in the Si-doped thin film samples studied in Ref. \onlinecite{korhonen2015} are somewhat further away from the bulk point than the $V_{\mathrm{Ga1}}$ $V_{\mathrm{Ga2}}$, while the Sn-doped and undoped thin film samples (Ref.~\onlinecite{spie}) are all very close to the bulk point. In those reports this behavior of the data is associated to different levels of hydrogenation of the samples due to different precursors in the synthesis. Our present results on the split Ga vacancies bring more detail to this interpretation. We suggest that in the Sn-doped and undoped \gao\ thin films the positron data are dominated by (hydrogenated) split Ga vacancies, in line with the orientation dependence of the signals seen in Ref.~\onlinecite{spie} and the Sn-type split Ga vacancies observed in Ref.~\onlinecite{stem}, where Sn substitutes for the Ga atom relaxing to the interstitial position. This relaxation does not happen when Si substitutes for Ga \cite{varley_proton,varley_splitvacancy}, and allows for the formation of regular Ga mono-vacancies observed in the experiments. This interpretation answers the question raised in those reports concerning the low formation energy of Ga vacancies - indeed experiments indicate that Ga vacancies are abundant in all \gao\ samples. Further, recent calculations \cite{varley_chapter} predict that the split Ga vacancies can accommodate three H atoms and become neutral and non-compensating. This provides a plausible solution to the electrical compensation question: split Ga vacancies are present at high concentrations in $n$-type material, but they only act as efficient compensating centers when not strongly hydrogenated.

\section{Summary}

We present a comprehensive theoretical study of positron states and positron-electron annihilation signals in the \gao\ lattice and the various Ga mono-vacancy defects that this lattice is able to host. We also performed systematic angle-resolved Doppler broadening experiments that confirm the findings obtained by the state-of-the-art calculations. We address the difficulties in studying defects with positron annihilation in \gao, caused by a combination of (i) signal anisotropy of unprecedented magnitude for 3D crystals and (ii) relatively small differences between signals originating from the \gao\ lattice and the various types of Ga mono-vacancy defects. In short, the positron signal anisotropy in \gao\ is larger than the difference between the lattice and vacancy signals, and the ($S,W$) parameters of \gao\ lattice and vacancies in various orientations overlap strongly. 

The unusual symmetry properties of the \gao\ lattice are at the origin of the anisotropy in several ways. First, the delocalized positron state in the \gao\ lattice is found to form one-dimensional tubes along the [010] lattice direction, in contrast to three-dimensionally delocalized positron states in more typical crystal structures with cubic or hexagonal symmetry. This leads to 5-10-fold differences of the ($S,W$) parameters measured along different lattice orientations compared to those found in, for example, Si or ZnO. Second, energetically favorable configurations of Ga mono-vacancies involve strong relaxations leading to highly non-symmetric split Ga vacancy configurations \cite{stem,varley_splitvacancy,varley_proton,weiser}. The localized positron state is highly non-spherical (anisotropic) in these defects, accompanied by an increase of the magnitude of the anisotropy by a factor of up to 3, in contrast to the typical mono-vacancy defects in other semiconductor crystals which exhibit at most the same barely noticeable magnitude of anisotropy as the lattice. Finally, the particularly low formation enthalpies \cite{varley_splitvacancy,varley_proton} of all of the Ga mono-vacancy defects in \gao\ lead to the situation where all samples appear to contain high concentrations ($>$\SI{1e18}{\centi\meter^{-3}}) of split Ga vacancies. This causes challenges for positron annihilation spectroscopies that are strongly comparative in nature and typically require a well-defined reference sample, preferably one that does not show positron trapping at any vacancy type defects, for detailed defect identification.

The colossal anisotropy of the Doppler broadening parameters allows for a different approach in defect identification in \gao. We demonstrate that by performing experiments in more than one direction, positron states can be differentiated thanks the differences in the nature of the signal anisotropy. Importantly, the signal anisotropy must be taken into account when performing Doppler broadening experiments in \gao, and the measurement direction in relation to the crystal orientation needs to be analyzed in detail when presenting experimental data. In future work on positron annihilation in \gao, it is imperative to pay close attention to the measurement directions and give a detailed account of the measurement geometry.

\begin{acknowledgments}

We wish to thank Mr. Vitomir Sever and Ms. Daria Kriukova for technical assistance in the angle-resolved Doppler experiments. We acknowledge the computational resources provided by CSC (Finnish IT Centre for Science). This work was partially supported by the Academy of Finland grants Nr 285809, 315082 and 319178. A. Karjalainen wishes to thank the Magnus Ehrnrooth foundation for financial support. This work was partially performed under the auspices of the U.S. DOE by Lawrence Livermore National Laboratory under contract DE-AC52-07NA27344, and supported by the Critical Materials Institute, an Energy Innovation Hub funded by the U.S. DOE, Office of Energy Efficiency and Renewable Energy, Advanced Manufacturing Office.
\end{acknowledgments}

\clearpage
\appendix

\section{Calculated ratio curves}
\label{appendix}

\begin{figure}[htb]
  \begin{center}
    \includegraphics{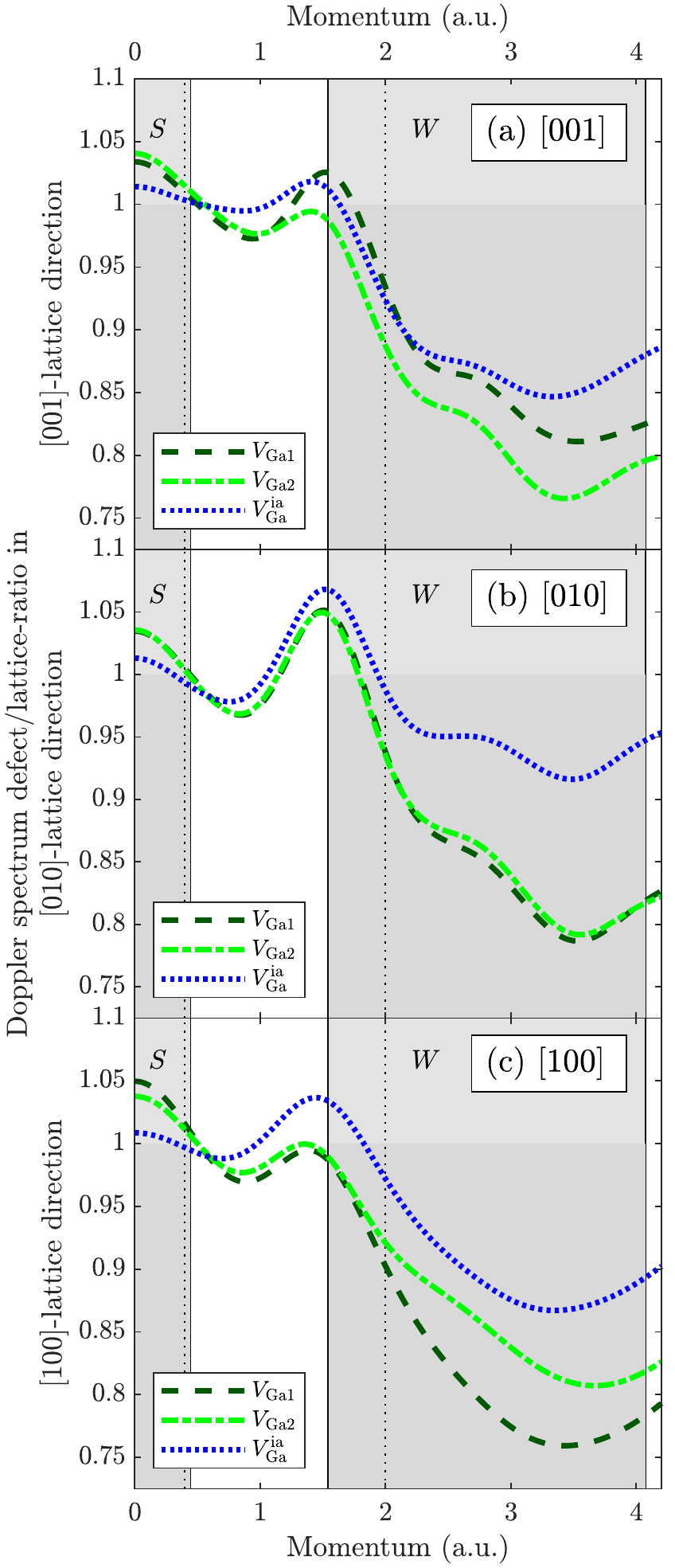} 
    \caption{Vacancy/lattice-ratios of defects \vgai, \vgaii\ and \via.}        
    \label{f:defect_ratios1}
  \end{center}
  \end{figure}

\begin{figure}[htb]
 \begin{center} 
 \includegraphics{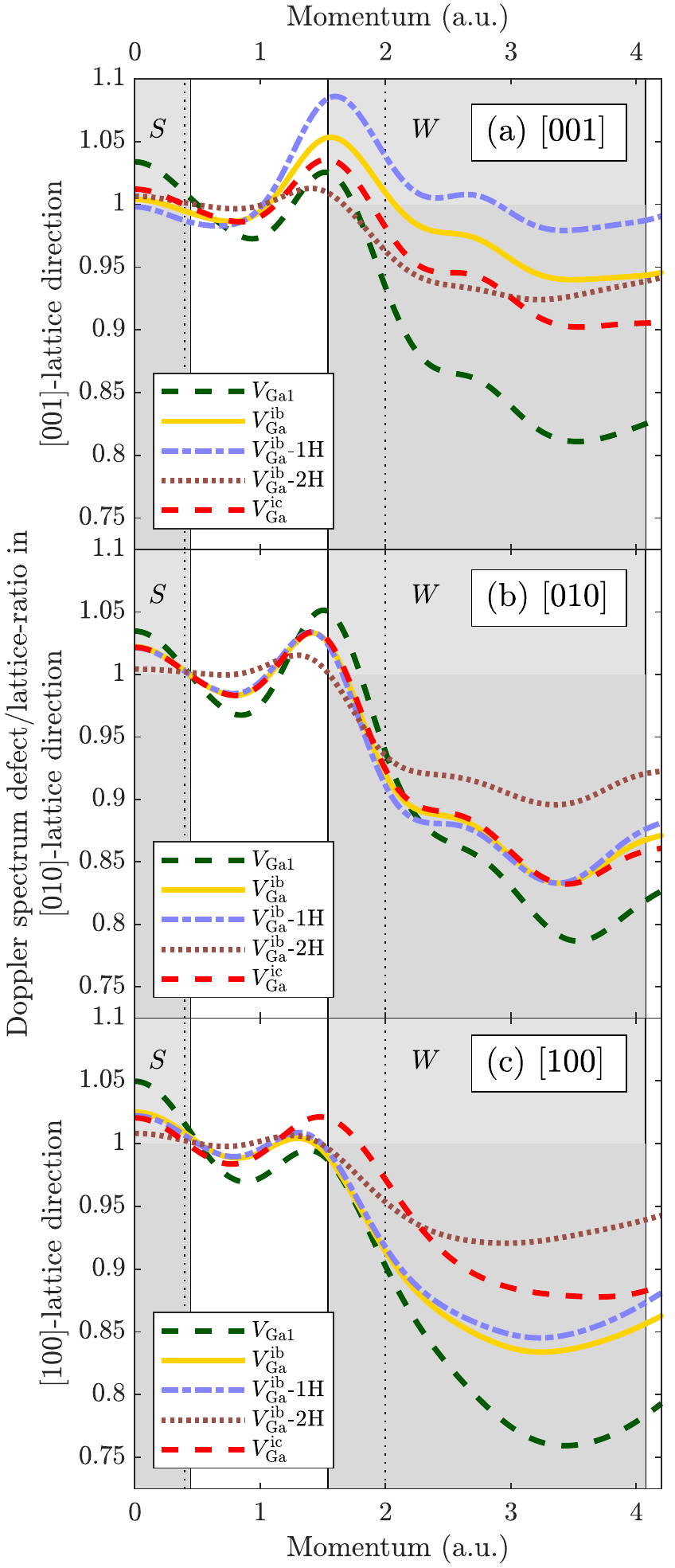}   
  \caption{Vacancy/lattice-ratios of defects \vgai, \vib~, \vib-1H, \vib-2H  and \vic.}
 \label{f:defect_ratios2}
  \end{center}
\end{figure}

\begin{figure}[htb]
  \begin{center}
    \includegraphics{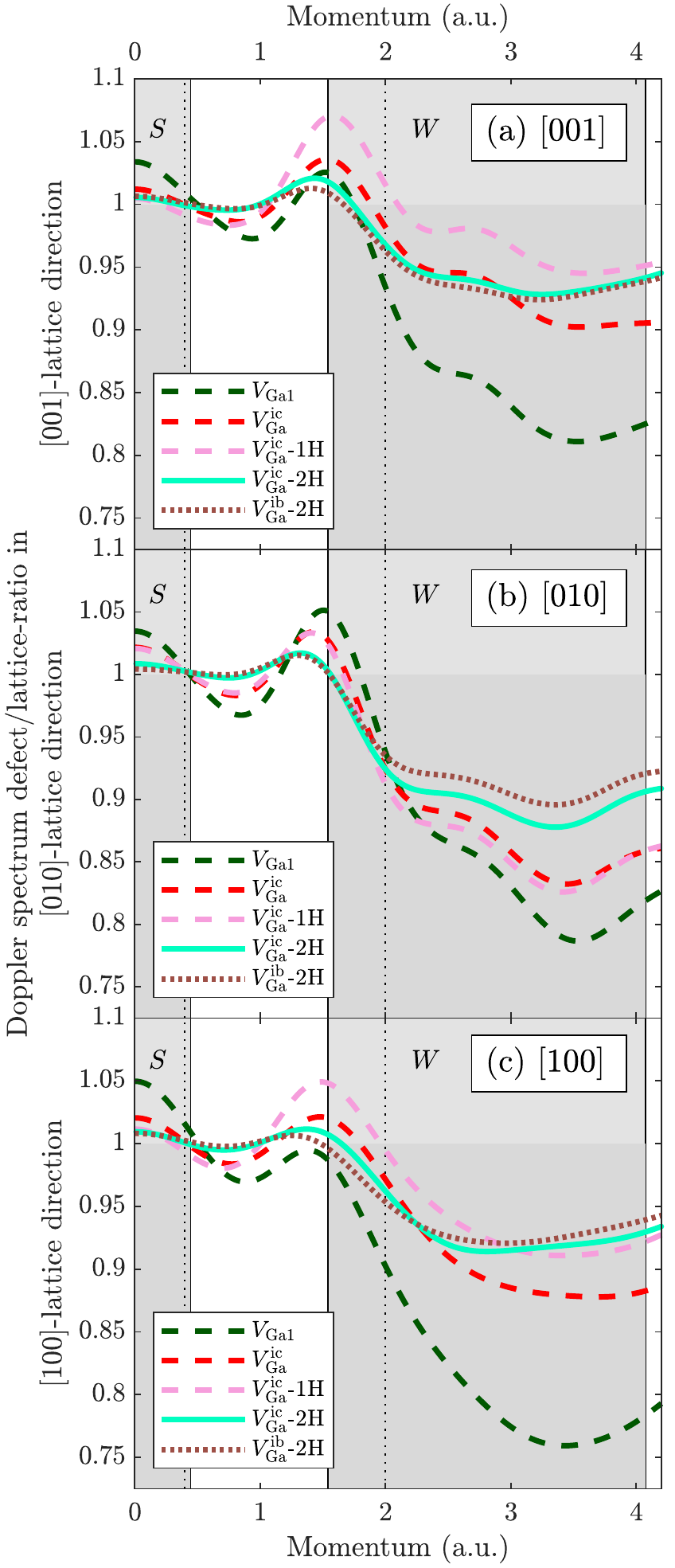}
  \end{center}
  \caption{Vacancy/lattice-ratios of defects \vgai, \vic~, \vic-1H, \vic-2H  and \vib-2H.}
    \label{f:defect_ratios3}
\end{figure}

We show for the benefit of future work the defect-lattice ratio curves for all the defects considered in this work in Figs. \ref{f:defect_ratios1}, \ref{f:defect_ratios2} and \ref{f:defect_ratios3}. The ratio curves are normalized by the lattice Doppler spectrum in each of the three directions ([001], [010] and [100]). Interestingly, the ratio curves in the [001] direction (for all defects) appear the most ordinary in the sense that their shapes remind 
those obtained for various types of defects in GaN \cite{HautakangasPRB2006,TuomistoPRL2017} and ZnO \cite{ZubiagaPRB2008,JohansenPRB2011}. In this direction, the shoulder-like feature around 1.5 a.u. is the least pronounced for most of the defects. Note that this shoulder-like feature, that in some cases resembles a peak, is typical of metal-oxide lattices where the atomic fraction of oxygen is higher than 50 \% \cite{MakkonenJPCMoxides}. The origin of this shoulder is in the strong contribution from O 2p electrons in this momentum range. At higher momenta, Ga 3d electrons dominate the annihilations. The strength of this feature could in principle be used to distinguish between the different defects in measurements performed in a given direction, in particular in cases where maximum value of the ratio switches from below 1 to above 1 (for example, regular mono-vacancy vs. split Ga vacancy). In practice, however, the reliability of this identification criterion is hard to assess as this is the momentum range where the functional shape of the Doppler spectrum changes from Gaussian to (sub-)exponential and the detector resolution plays a significant role \cite{LinezPRB2016}. Comparing the differences in the anisotropic behavior of the defects remains more reliable an approach in \gao.

The shoulder/peak feature at around 1.5 a.u. is at the origin of the $S$ parameter decreasing below that of the \gao~ lattice for some of the defects. In addition, it dominates the signal intensity of the $W$ parameter region as seen from the grey shaded areas in Figs. \ref{f:defect_ratios1}, \ref{f:defect_ratios2} and \ref{f:defect_ratios3}. This suggests that the difficulties with the overlap of the lattice and vacancy signals could be mitigated to some extent by modifying the ($S,W$) parameter integration windows for improved differentiation, at the expense of statistical accuracy. In Appendix B, we consider alternative ($S,W$) parameter integration windows shown with the dotted lines in the ratio curve plots. The windows are changed from [0, 0.45] a.u. to [0, 0.40] a.u. for the $S$ parameter and the lower limit of the $W$ windows is shifted from 1.54 a.u. to 2.0 a.u. The upper limit of $W$ parameter was not changed as the momentum distribution intensity is almost negligible already at 4 a.u.

\section{Additional $(S,W)$ figures}

\begin{figure}[htb]
  \begin{center}
    \includegraphics{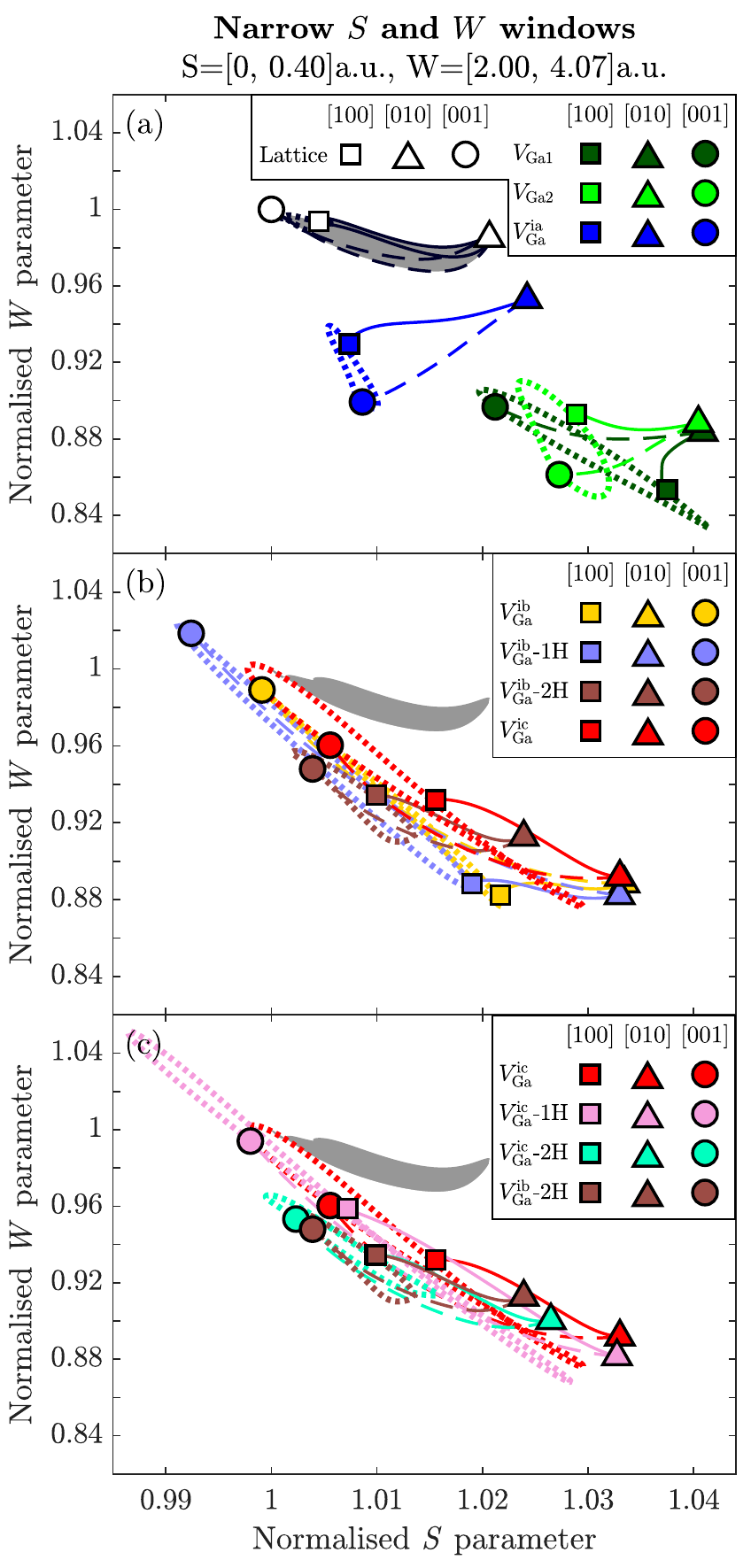}
  \caption{The ($S,W$) parameters of the lattice and vacancies calculated with alternative ($S,W$) parameter windows. The ($S,W$) parameters are normalized to the $\beta$-Ga$_2$O$_3$ lattice in the [001] lattice direction. The grey shadow illustrates the ($S,W$) parameter of the $\beta$-Ga$_2$O$_3$ lattice. The figure uses the notation introduced in Fig. \ref{f:bulk_anisotropy}. }
\label{f:narrow_windows}
  \end{center}
  \end{figure}

Fig. \ref{f:narrow_windows} shows the ($S,W$) parameters analyzed with the narrow $S$ and $W$ windows, normalized to $\beta$-Ga$_2$O$_3$ lattice in the [001] lattice direction. The effect of changing the $S$ parameter window on the relative $S$ parameters is minimal. The removal of the shoulder effect from the $W$ parameter decreases the $W$ parameter of vacancy defects and improves the overlap situation, but not dramatically. With the $W$ parameter windows starting from 2.00 a.u., the \gao~ lattice produces the highest $W$ parameter in the [100] and [010] lattice directions, providing additional means to distinguish between the lattice and defects. However, this narrowed $W$ parameter window reduces the statistical weight of the $W$ parameter by a factor of 3 to only about 2 \% of the whole 511-keV annihilation peak, making analysis less accurate. Generally, changes in the $S$ parameter are more reliably monitored. Nevertheless, the future process of identifying a "defect-free" reference \gao\ sample will benefit from examining alternative $W$ parameter windows to differentiate between defect and lattice signals.

\begin{figure}[htbp]
\begin{center}

\includegraphics{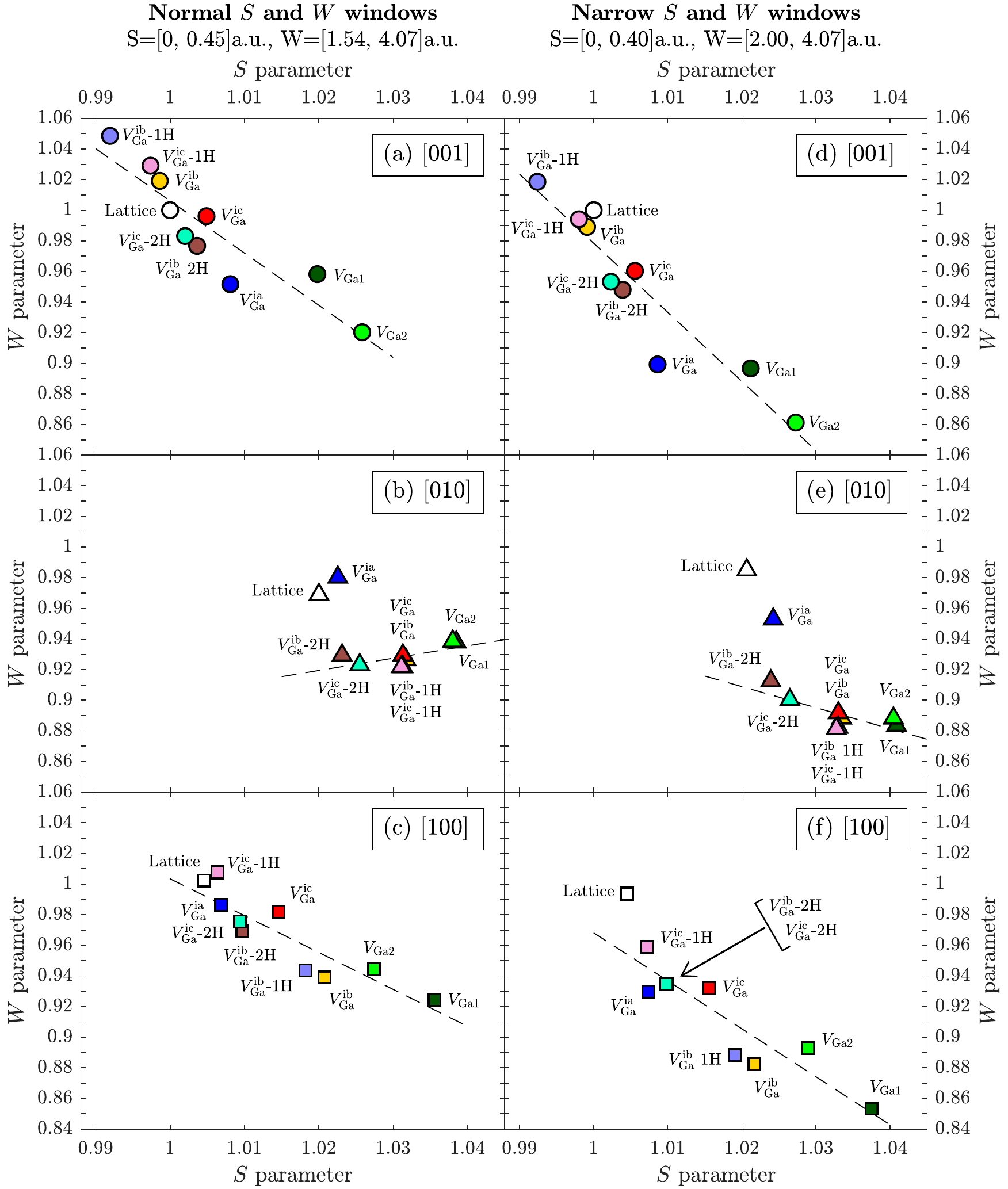}

\caption{Calculated ($S,W$) parameters in the $\beta$-Ga$_2$O$_3$ lattice and the vacancy defects sorted in (a)\&(d) [001], (b)\&(d) [010] and (c)\&(f) [100] lattice directions to illustrate the different behaviour in slow positron beam experiments in selected directions. Figures (a)-(c) show the normal ($S,W$) parameters used in this work. Figures (d)-(f) show ($S,W$) parameters with narrower integration windows. The ($S,W$) parameters are normalized to the ($S,W$) parameters of $\beta$-Ga$_2$O$_3$ lattice in the [001] lattice direction. The dashed lines are drawn to guide the eye. }
\label{f:sw_directions}
\end{center}
\end{figure}

For the benefit of future defect studies in \gao, Figure \ref{f:sw_directions} illustrates how the normalized ($S,W$) of $\beta$-Ga$_2$O$_3$ lattice and defects behave in the [001], [010] and [100] lattice directions for both sets of integration windows. The calculated ($S,W$) parameters in those lattice directions have different overall trends and magnitudes of differences between defects vary. The regular mono-vacancies $V_{\mathrm{Ga1}}$ and $V_{\mathrm{Ga2}}$ show the largest $S$ parameter in all lattice directions the relative positions of the ($S,W$) parameters of the other defects change from one direction to another. 

The trends of the ($S,W$) parameters in the [001] and [100] lattice directions are rather similar and follow a diagonal line. As discussed in Section \ref{lattice}, the details of the relative ($S,W$) parameter positions in these two directions may depend on the exact atomic positions in the unit cell. There is, however, an important aspect that can be seen in the [001] data in Fig.~\ref{f:sw_directions}a and is even more pronounced in Fig.~\ref{f:dft_sw_tiled}: some of the defects, such as $V_{\mathrm{Ga}}^{\mathrm{ib}}$-H and $V_{\mathrm{Ga}}^{\mathrm{ic}}$-H, are characterised by a shift of the ($S,W$) parameters towards the upper left corner from the \gao\ lattice point, even when measured in the same direction. This needs to be taken into account when attempting an interpretation of experimental positron data in \gao, as usually the lattice produces the lowest $S$ and highest $W$ parameters. The reason for this can be seen in the ratio curves. In the [010] direction, the ($S,W$) parameters of the defects behave completely differently and set on an nearly horizontal line (excluding $V_{\mathrm{Ga}}^{\mathrm{ia}}$) where the $S$ parameter roughly follows the lifetime and the size of the open volume. Importantly, the ($S,W$) parameters of the defects are no longer aligned with those of the lattice, providing additional means for distinguishing defect-related phenomena from anisotropy-related phenomena in the experimental data.

The main effect of employing narrower $S$ and $W$ parameter windows in Fig. \ref{f:sw_directions} is in the widening the span of the $W$ parameters thanks to the non-inclusion of the high-intensity shoulder-like feature (see the ratio curves of the previous section). At first sight the changes are not dramatic, but by comparing the behavior in the [010] direction to any of the two other directions a feature is found that may allow, for example, to distinguish between regular and split Ga vacancies in experiments where the full 3D anisotropy is not meusured. The regular Ga vacancies have a lower (or at most similar) $W$ parameter than the split Ga vacancies for all three measurement directions when narrow integration windows are employed, while it is higher with normal integration windows in the [010] direction. In addition, when going from normal to narrow integration windows the differences become larger when they are in the samde direction originally. Monitoring these types of trends in experimental data may bring additional insight to defect idenfitication.

\onecolumngrid

\end{document}